\documentclass[sigconf, nonacm]{acmart}

\AtBeginDocument{%
  }

\setcopyright{acmlicensed}
\copyrightyear{2026}
\acmYear{2026}
\acmDOI{10.1145/3797867.3829049}

\acmConference[ASSETS '26]{The 28th International ACM SIGACCESS Conference on Computers and Accessibility}{October 25--28,
  2026}{Porto, Portugal}
\acmISBN{979-8-4007-2521-0/2026/10}

\usepackage[table,xcdraw]{xcolor}
\usepackage{xcolor}
\usepackage{multirow}
\usepackage{enumitem}
\usepackage{soul}
\usepackage{xcolor}
\usepackage{ifthen}
\usepackage{tcolorbox}
\usepackage{graphicx}
\usepackage{subcaption} % For subfigures if you want multiple plots together

\newboolean{showchanges}
\setboolean{showchanges}{true}

\definecolor{revisionblue}{RGB}{220, 235, 255} 

\newcommand{\rev}[1]{%
  \ifthenelse{\boolean{showchanges}}%
    {\sethlcolor{revisionblue}\hl{#1}}%
    {#1}%
}
\newcommand{\revsection}[1]{%
  \ifthenelse{\boolean{showchanges}}%
    {\begin{tcolorbox}[
      colback=revisionblue,  
      colframe=revisionblue,  
      boxrule=0pt,            
      arc=0pt, 
      left=2pt, 
      right=2pt, 
      top=2pt, 
      bottom=2pt
    ]#1\end{tcolorbox}}%
    {#1}%
}

\newif\ifshowchanges 
\showchangesfalse % Uncomment this line for the clean version and comment the line above
\newcommand{\changed}[1]{%
    \ifshowchanges 
        \textcolor{blue}{#1}%
    \else 
        #1%
    \fi
}

\newif\ifshowchangescamera
\newcommand{\changedcamera}[1]{%
    \ifshowchangescamera 
        \textcolor{purple}{#1}%
    \else 
        #1%
    \fi
}

\copyrightyear{2026}
\acmYear{2026}
\setcopyright{cc}
\setcctype{by}
\acmConference[ASSETS '26]{The 28th International ACM SIGACCESS Conference on Computers and Accessibility}{October 25--28, 2026}{Vila Nova de Gaia, Portugal}
\acmBooktitle{The 28th International ACM SIGACCESS Conference on Computers and Accessibility (ASSETS '26), October 25--28, 2026, Vila Nova de Gaia, Portugal}
\acmDOI{10.1145/3797867.3829049}
\acmISBN{979-8-4007-2521-0/2026/10}

\begin{document}

%% Commands
\newcommand{\system}{ColorA11Y}
% Color-A11Y, DesignPilot, CABA, ADA, 
\newcommand{\todo}[1]{\textcolor{red}{TODO: #1}}

%%
%% Title
\title{\system: Enhancing Creative Design Workflows with Just-in-Time Color Accessibility Recommendations}



\author{Alexa Siu}
\email{asiu@adobe.com}
\affiliation{
  \department{Adobe Research} 
  \city{San Jose}
  \state{CA}
  \country{USA}
}

\author{Rajiv Jain}
\email{rajijain@adobe.com}
\affiliation{
  \department{Adobe Research} 
  \city{College Park}
  \state{MD}
  \country{USA}
}

\author{Abhinav Kannan}\authornote{Work completed during an internship with Adobe Research, College Park, MD, USA.}
\email{abhinavk@umd.edu}
\affiliation{%
  \department{College of Information} \institution{University of Maryland}
  \city{College Park}
  \state{MD}
  \country{USA}
}
% \additionalaffiliation{%
%   \institution{Adobe Research, College Park, MD, USA}
%   \city{College Park}
%   \state{MD}

\author{Jose Echevarria}
\email{echevarr@adobe.com}
\affiliation{
  \department{Adobe Research}
  \city{New York}
  \state{NY}
  \country{USA}
}

\author{Mary Ann (MJ) Jawili}
\email{mj@mjawili.com}
\affiliation{
  \department{ServiceNow Inc} 
  \city{San Francisco}
  \state{CA}
  \country{USA}
}

\author{Yalpi Shiva Prasad}
\email{yshivu@gmail.com}
\affiliation{
  \department{RVCE} 
  \city{Bengaluru}
  \state{Karnataka}
  \country{India}
}

\author{Rick Treitman}
\email{rick@treitman.com}
\affiliation{
  \department{Adobe Inc} 
  \city{Newton}
  \state{MA}
  \country{USA}
}

\author{Garreth W. Tigwell}
\email{garreth.w.tigwell@rit.edu}
\affiliation{%
  \department{School of Information} \institution{Rochester Institute of Technology}
  \city{Rochester}
  \state{NY}
  \country{USA}
}

\author{Jonathan Lazar}
\email{jlazar@umd.edu}
\affiliation{%
  \department{Maryland Initiative for Digital Accessibility, College of Information} \institution{University of Maryland}
  \city{College Park}
  \state{MD}
  \country{USA}
}
\additionalaffiliation{%
  \institution{Engineering Design Centre, University of Cambridge, Cambridge, UK}
  \city{Cambridge}
  \country{United Kingdom}
}
\additionalaffiliation{%
  \institution{Faculty of Information, University of Toronto, Toronto, Canada}
  \city{Toronto}
  \country{Canada}
}

\renewcommand{\shortauthors}{Siu et al.}

%TC:ignore
%%% word count: 143 / 150
\begin{abstract}
Effective color contrast in visual design is essential for content accessibility. While existing tools can identify contrast issues, they often operate in isolation from design workflows or are used as an afterthought. We present \system, a system that supports designers in creating accessible content by providing just-in-time feedback and actionable recommendations throughout the authoring process to meet accessibility color contrast guidelines. 
Our system analyzes the visual properties of text and background elements and offers recommended changes, including text color adjustments, background modifications, and opacity changes. Through two user studies, we evaluate \system's effectiveness. A user preference study (n=40) revealed varying effectiveness of different recommendations based on design context, while a qualitative study with designers (n=8) indicated a more seamless workflow experience \changed{ in comparison to a baseline using a color contrast checker.}
% a more seamless workflow experience. 
This work advances a \emph{born-accessible} approach to design, where accessibility considerations are seamlessly integrated into the creative process rather than treated as an afterthought.
\end{abstract}


%%
%% Keywords.

%%
%% Teaser
\begin{teaserfigure}
  \includegraphics[width=\textwidth]{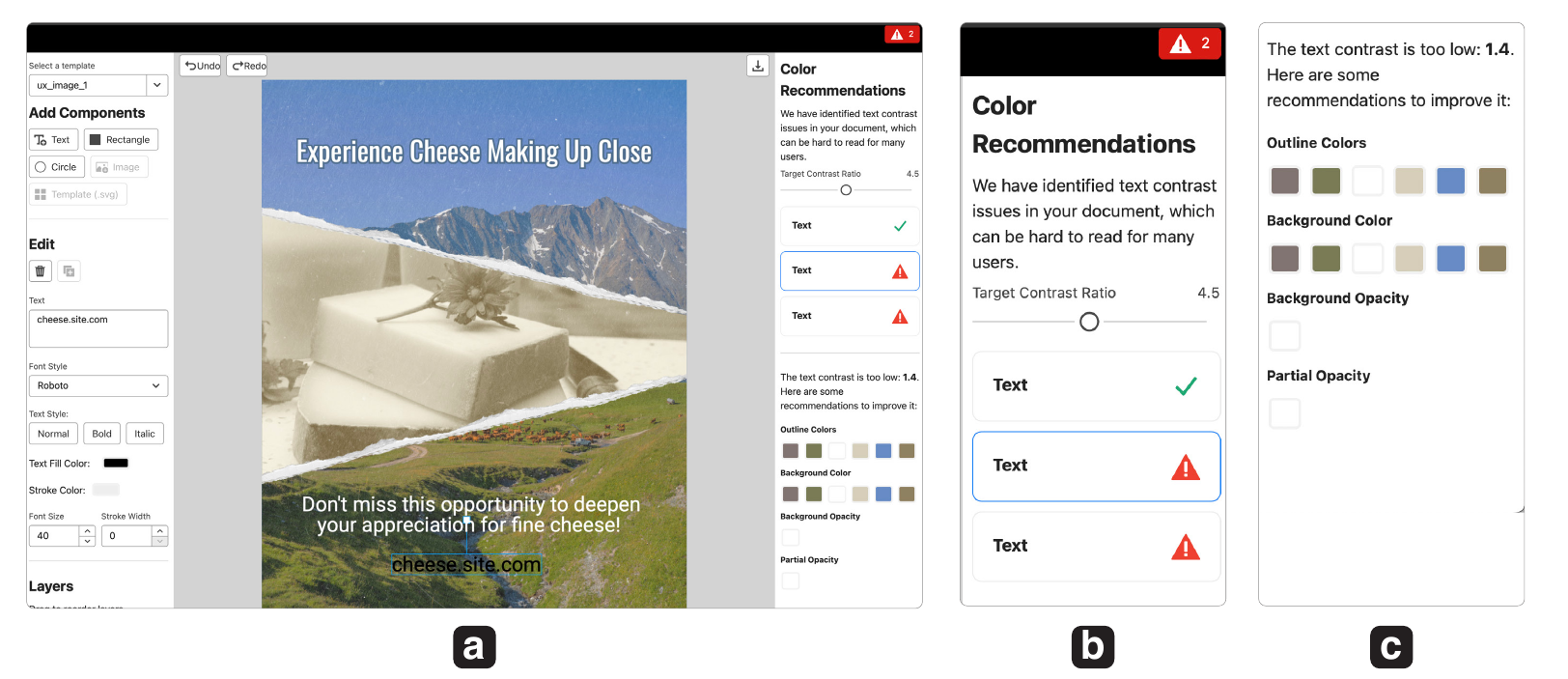}
  \caption{\system{} (a) supports designers in creating accessible content by providing just-in-time feedback (b), and actionable recommendations (c) throughout the authoring process with settings to meet accessibility color contrast guidelines.}
  \Description{Three panels demonstrating the ColorA11Y tool for accessible design. Panel (a) displays the main interface where a user is editing a promotional image for cheese making, with text editing controls on the left and the main canvas editor in the center. Panel (b) shows a feedback window indicating text contrast issues with a target contrast ratio of 4.5. The element with text contrast issues are highlighted. Panel (c) shows specific color recommendations to improve contrast, including suggested outline colors, background colors, and opacity settings. The interface shows how ColorA11Y provides real-time accessibility feedback and actionable recommendations to help designers meet color contrast guidelines while authoring content.}
  \label{fig:teaser}
\end{teaserfigure}


\maketitle

\section{Introduction}

Color is an important element in the design of digital assets (e.g., posters, flyers, social media posts). Color choices influence aesthetics, readability, and accessibility~\cite{mithun2019impact}. When used thoughtfully, color choices can enhance readability and visual appeal, but poor decisions can create significant barriers to access for people with visual impairments~\cite{WCAG_Contrast_Minimum}, dyslexia~\cite{rello2017good}, situational impairments~\cite{tigwell2018its}, or even increase viewers' fatigue~\cite{xie2021study}. One of the most widely used accessibility standards is the Web Content Accessibility Guidelines (WCAG),\footnote{https://www.w3.org/WAI/standards-guidelines/wcag/} which offers guidelines for ensuring content accessibility, including color contrast recommendations. \changed{Providing sufficient text color contrast is encouraged by WCAG because it enables improved readability for all individuals, accommodating for low vision and color vision deficiencies~\cite{WCAG_Contrast_Minimum}}. WCAG is widely adopted and is used by accessibility laws such as the revised Section 508 in the United States~\cite{section508_update_2017} and even other accessibility standards such as the European Union's EN 301 549~\cite{wcag_eaa_compliance_2024}. Yet designers often lack clarity or sufficient awareness of accessibility practices to know how to meaningfully integrate these principles into design workflows~\cite{andrew2022accessible,lewthwaite2023workplace}. Designers may also face the challenge of balancing accessibility requirements with the aesthetic or visual appeal of their creations, especially when dealing with complex content such as gradients, images, and textured backgrounds. In this work, we explore how to support designers in creating accessible content by providing just-in-time feedback and actionable recommendations during the authoring process to meet accessibility color contrast guidelines.

Existing tools provide some support for improving color accessibility but often focus narrowly on specific tasks, such as selecting compliant color palettes or checking luminance contrast~\cite{Coolors, iWantHue, colormaker, ColorBrewer2}. While useful, these tools frequently lack integration into broader design workflows, leaving users to bridge the gap between identifying issues and knowing how to implement proper fixes~\cite{hadadi2021adee, kokate2022exploring}. Additionally, these tools frequently overlook complex color scenarios (focusing on solid colors), often failing to consider the intricacies of design elements common in modern real-world visual content~\cite{almeida2020analysis}.
% such as marketing materials, presentations, and illustrations ~\cite{almeida2020analysis}. 
Almeida and Duarte report that many popular contrast checker solutions misrepresent non-solid backgrounds and yield inaccurate pass/fail WCAG conditions~\cite{almeida2020analysis}. This disconnect between identifying and addressing accessibility issues often increases the cognitive load on designers, making it difficult to balance aesthetic goals with accessibility requirements. Without actionable, workflow-integrated guidance, accessibility becomes a secondary consideration that is often addressed only after the creative process has concluded.

To address these challenges, we introduce \system, a system that seamlessly integrates color accessibility guidance into the document authoring workflow (Figure~\ref{fig:teaser}). 
Our approach aims to advocate for a shift toward a \emph{born-accessible} approach~\cite{lazar2023framework, lazar2026bornaccessible}, introducing proactive and context-aware recommendations that aim to help users maintain accessibility requirements throughout the design process rather than as a final checkpoint.  
While the user is editing their content, \system{} provides just-in-time feedback, alerting the user of accessibility issues and actionable recommendations that help the user address the issues. The system analyzes the visual properties of text and background elements to detect potential accessibility issues (Figure~\ref{fig:teaser}b) and provides various recommendations for improving color contrast (Figure~\ref{fig:teaser}c) even in cases of complex backgrounds. These recommendations include adjusting text color, adding backgrounds or outlines, and modifying opacity levels---all while maintaining WCAG contrast requirements for standard text.

Through two user studies, we evaluate the proposed tool's impact. The first study (Section~\ref{sec:study_1_survey}) explores user preferences for accessible recommendations that balance readability and visual appeal (n=40). The effectiveness of different recommendations varied significantly based on the underlying design context (e.g., image, texture, or gradient background) and informs the use of different recommendations. 
The second study (Section~\ref{sec:study_2_experience}) examines how \system{} integrates into authoring workflows, assessing its ability to raise awareness and improve accessibility practices compared to a baseline tool \changed{using a color contrast checker} that does not integrate accessibility during authoring. Results from interviews with 8 non-professional designers demonstrate that \system{} improved users' workflow when creating accessible designs, with participants \changedcamera{subjectively} rating the tool higher for ease of use (4.5/5), time efficiency (4.8/5), and support (4.5/5) \changed{as well as generating designs with fewer text color contrast issues}. Importantly, participants reported that the just-in-time alerts and recommendations helped them learn about accessibility considerations while maintaining their creative workflow. Participants perceived the alerts as helpful nudges rather than interruptions to their work, noting how they visibly led to improvements in their designs. The actionable recommendations helped users bridge the gap between identifying issues and implementing solutions that worked for their design. 

%Our approach aims to advocate for a shift toward a "born-accessible" approach, embedding accessibility considerations seamlessly as part of the authoring process. Rather than treating accessibility as a final checkpoint, a born accessible approach integrates color accessibility checking as a core part of the design workflow, empowering users to make informed, inclusive decisions without compromising creativity. While our broader vision encompasses support for all aspects of accessibility, in this work, we focus specifically on color accessibility as a step towards born-accessible creation.

To summarize, our contributions are as follows:
\begin{enumerate}
    \item \system, a real-time color recommendation system that integrates into document authoring workflows (i.e., a born-accessible approach) to assist users when authoring documents with complex backgrounds.
    \item Empirical findings \changed{that} show that designers’ choices for text contrast depend on the document content and their individual preferences. 
    \item  \changedcamera{Evidence that a \emph{born-accessible} approach to color accessibility} can be compatible with authoring workflows for designers, where nudges toward better contrast outcomes without overriding designer choices demonstrate how accessibility support can be embedded in creative tools while maintaining designer autonomy.
\end{enumerate}

\section{
\changed{Background}
}
\label{sec:related_work}

\subsection{Importance of Color Accessibility}
To consume visual content, sufficient contrast in luminance between text and its background is critical for anyone with or without a disability. There are two main reasons why it is important to make appropriate use of colors for accessible designs. First, vision disabilities can significantly impact sensitivity to contrast and require more contrast between text and its background. Approximately 300 million people worldwide have color vision deficiency (CVD)~\cite{colourblindawareness}, with CVD being more prevalent in the male population~\cite{birch2001diagnosis}. Second, people without CVD can also experience challenges with distinguishing colors that do not have sufficient contrast due to situational impairments. For example, this might be caused by viewing content on displays under bright sunlight and the colors become more washed out~\cite{liu2014effect,tigwell2018its}, changes in display characteristics or user-worn accessories like sunglasses~\cite{flatla2012situation,tigwell2018its}, or when designs are printed in grayscale~\cite{menzies2022author}.

When creating digital assets, there are several guidelines that designers can follow to ensure colors are more accessible~\cite{section508_update_2017, apple_accessibility_vision, material_design_color_contrast, microsoft_inclusive_design}. WCAG is one of the most widely used accessibility standards, which offers guidelines for ensuring content accessibility, including color contrast recommendations. Although WCAG's name suggests web accessibility, the color contrast recommendations have been used to inform accessible design in other areas, such as mobile app design~\cite{apple_accessibility_vision,material_design_color_contrast}. However, studies have demonstrated that despite the existence of the guidelines, designers do not follow them, resulting in color accessibility issues~\cite{kuzma2010accessibility,patra2014quantitative,pillai2022websites,webaim2021the}. An alternative approach to improve color accessibility is for the design tools to guide the designer during use.

Beyond standard WCAG color contrast ratio calculations, there are alternative algorithms for evaluating color contrast. Prior research has noted limitations with WCAG color contrast calculations when applied to complex backgrounds~\cite{almeida2020analysis,sharma_2024,oconnor2019orange}, and alternative calculations have been proposed to address these issues, such as the Accessible Perceptual Contrast Algorithm~\cite{myndex2022apcaeasyintro,moreno_designing_2024}. While \system{} follows the more popular WCAG, our tool is robust and can easily use other color contrast algorithms instead. Even within WCAG, \system{} allows users to adjust the target contrast ratio.

\subsection{Design Tools to Support Color Accessibility}

Design tools that support accessibility, including color design, can be broadly categorized into tools that integrate into the design process (e.g.,~\cite{hadadi2021adee,tigwell2017ace}) and tools that provide post hoc evaluations (e.g.,~\cite{vienot1999digital,reinecke2016enabling,padure2020comparing}). Our research is focused on the former.

One example tool from prior work is the Accessible Colour Evaluator (ACE)~\cite{tigwell2017ace}, which supports designers in exploring a color palette that meets their aesthetic design requirements while being informed which color pairs in the palette can be used together for accessibility. However, ACE is a standalone website that does not integrate with a designer's preferred design tool. Prototyping tools and other design platforms rarely integrate sufficient accessibility features, with most relying on third-party plugins~\cite{kokate2022exploring} (e.g., Adee~\cite{hadadi2021adee}). However, plugins can still be limited in features~\cite{frazao2020comparing}, and there may be quality control issues compared to companies building accessibility directly into the design tool~\cite{kokate2022exploring}. More recently, Figma\footnote{http://figma.com} (a popular canvas design tool) released an in-built color contrast checker. When enabled, this tool checks color contrast of a selected element on the canvas against WCAG AA and AAA standards. While this feature signals whether the contrast ratio is met, it does not include recommendations that assist the designer in improving their design.

To better assist designers' choices, there have been a number of accessibility-focused tools that guide color palette selection for UX design~\cite{Coolors}, with some focusing further---color maps for map data visualization~\cite{iWantHue, colormaker, ColorBrewer2}. While all of these tools are standalone, they have taken significant steps toward including color-safe features for people with CVD. Users can provide color preferences as input, customize and filter their selection, and obtain an exported set of optimized color choices. This can then be imported by designers into their preferred design environment.

However, accessibility tools that are separate from the design tool can be “an overhead for switching back and forth”~\cite{hegemann2024palette}. Tigwell et al. found that only about two-thirds of designers are aware of accessible color guidelines, but only about one-third actually apply them~\cite{tigwell2017ace}. Our work takes a step in addressing these gaps of overhead and explanatory suggestiveness by providing context-driven color choices during the design process. Rather than designers choosing colors to meet accessibility standards/guidelines~\cite{hegemann2024palette} alone, 
\changedcamera{our work aims to utilize them as an opportunity for better design choices and broader audience inclusion.}
% our work aims to utilize them as an opportunity for UI innovation, better design choices, and broader audience inclusion.

\subsection{Born Accessible Design}

Technologies and content are often designed in a way that is inaccessible for people with disabilities~\cite{lazar2015ensuring}. Either remediation for accessibility is left until the end of the design process, or remediation for accessibility is never done. If remediation is never done, this puts the onus on people with disabilities to advocate and file complaints to get the content fixed. Even if the content is remediated, this leads to a situation where, for a period of time, people with disabilities don't have equal access while other people without disabilities do have access~\cite{wentz2011retrofitting}. This remediation approach is neither cost-effective or equitable, and more often leads to less innovative design~\cite{accessibilityFirst}.

A better approach for both software and content design is shifting towards \emph{born-accessible design}, where accessibility is a primary design goal, people with disabilities are included in the process from the beginning, and there is never a gap in access or people with disabilities needing to advocate for fixes~\cite{lazar2023framework, lazar2026bornaccessible}. The concepts of born-accessible design are now starting to be included in government policies. For instance, in the new regulation for state and local government in the US under Title II of the Americans with Disabilities Act, the remediation approach is no longer allowed, as public institutions such as universities are required to \textit{"to use the two- or three-year compliance time frame to prepare to make course content accessible proactively, instead of having to scramble to remediate content reactively."}~\cite{doitNDmaryland}. 
% The new rule further states: \textit{"The Department believes the better approach is to…avoid the need for public educational institutions to make content accessible on an expedited time frame on the back end, and to instead require public entities to treat course content like any other content covered by subpart H…[this] should generally obviate the need for students with disabilities to make individualized requests for course content that complies with WCAG 2.1 Level AA."}
% ~\cite{maryland_digital_2024}. 
The State of Maryland also now requires born-accessible design, not remediation, for web content.

Given that the legal requirements for content are moving in this direction and the established research on consequences in remediation approaches, we need design tools that support content creators in making accessible design choices while they are  creating the content. Such tools would allow authors to consider accessibility as part of their design flow, rather than in a separate tool, after the end of the design process. \system{} is a step in that direction.
\section{Design of \system}

In a typical design workflow, \changedcamera{accessibility is often a consideration only \textit{after} content creation is complete}~\cite{accessibilityFirst, hadadi2021adee} rather than being integrated and considered throughout the design process. 
% This remediation-focused approach results in inaccessible content~\cite{brinn2022framework}. 
%Rather than treating accessibility as a final checkpoint, our primary goal in this work was to explore a born-accessible approach~\cite{lazar2023framework} where accessibility is seamlessly integrated and considered throughout the design process. 
%This builds on prior work showing that integrated accessibility features in design tools can lead to more accessible outcomes and overall more elegant and effective solutions~\cite{hadadi2021adee, henry2007just}. 

In this section, we motivate the design of \system{\changed{, which aims to be embedded within the creative process~\cite{frich2019mapping}}}. \changedcamera{We first present a motivating scenario that illustrates the challenges we aim to address in a remediation workflow, then articulate the design goals for \system.}

\subsection{Motivating Scenario \& Challenges}

Consider Ben, a small business owner creating social media content for a product. He uses design tools to craft visually striking posts incorporating existing product photos, brand colors, and promotional text. Ben works on his design and runs an accessibility checker before publishing it. The tool alerts him to multiple color contrast violations in his designs. The white text he chose for product descriptions becomes nearly invisible against the busy product photos. The brand's signature color, while visually appealing, fails to meet WCAG contrast requirements when used for an important call-to-action text. Ben now goes back to the drawing board and attempts to remediate his design. To improve the call-to-action text, he decides to make a compromise and not use the brand's signature color. Pressed for time and unsure how to improve the contrast between the text and product photos and not wanting to change the photos, he leaves the description text unchanged without addressing the contrast issues.

This scenario illustrates several key challenges reflected in prior work:
\begin{enumerate}[label=C\arabic*]
    \item \textbf{Late-stage accessibility checks often lead to more costly redesigns and user effort}. Research shows that treating accessibility as a final checkpoint makes remediation expensive and disruptive, while incorporating it early reduces costs significantly~\cite{horton2024techbrief, w3c2014start}.
    \item \textbf{Users face cognitive load trying to balance aesthetic goals with accessibility requirements}. Designers struggle to simultaneously manage visual appeal and accessibility guidelines, particularly when selecting color palettes that must satisfy both aesthetic and contrast requirements~\cite{swallow2014speaking, tigwell2017ace}.
    \item \textbf{Current tools provide limited guidance on implementing accessibility fixes}~\cite{andrew2022accessible, tigwell2017ace, putnam2023could}. While tools can identify violations, they often fail to provide clear direction on solutions, leaving practitioners to figure out implementation on their own~\cite{andrew2022accessible, putnam2023could}.
    \item \textbf{The disconnect between identifying and addressing accessibility issues often leads to compromised solutions}~\cite{lazar2015ensuring}. Most accessibility tools require stepping outside primary design environments, creating workflow disruptions that reduce consistent usage~\cite{hadadi2021adee, kokate2022exploring, tigwell2017ace}.
\end{enumerate}

\subsection{Design Goals}
\label{sec:design_goals}
Drawing from both prior work and the challenges outlined, we articulate the following design goals by applying a born-accessible approach:
\begin{enumerate}[label=DG\arabic*]
    \item \textbf{Integrate accessibility into the authoring workflow.} Instead of doing an accessibility check as a final step, accessibility checks should be embedded directly into the content creation workflow. Research recommends accessibility features should be directly integrated to increase usage and reduce burden on the author~\cite{kokate2022exploring, henry2007just, accessibilityFirst}. \system{} implements this through just-in-time accessibility feedback while the user is editing content, thus encouraging continuous accessibility improvements throughout the authoring process.
    \item \textbf{Help users identify and resolve accessibility issues through actionable feedback.} Accessibility checks should not only identify issues but also provide clear and practical recommendations for resolution. Prior work has shown that simply identifying accessibility issues is insufficient; users also benefit from clear guidance on how to implement solutions~\cite{mehralian2024automated}. When \system{} identifies a color accessibility issue, it also suggests multiple contextual recommendations that improve the design. 
    \item \textbf{Build accessibility awareness through contextual explanations.} Research has shown that limited understanding of accessibility guidelines is a key barrier to creating accessible content~\cite{swallow2014speaking,tigwell2018designing, spyridonis2019serious, accessibilityFirst}. Rather than relying on separate documentation or training, our tool integrates educational content directly into the design workflow. By providing relevant explanations and guidance at the moment of need, we aim to help designers build lasting accessibility knowledge while completing their immediate tasks. 
    
    \item \textbf{Balance creative freedom with accessibility guidelines.} The tool should enable users to maintain their creative vision while incorporating accessibility guidelines. In addition to recommending text color changes, \system{} also provides a diverse set of design recommendations, including background colors, outlines, and partial opacity effects. Additionally, the recommendations aim to use colors in the same color family used in the design, which the user can further modify. \changed{Users can leverage the recommendations as a starting point and further adapt and edit to meet their design intent.}
    \changedcamera{We scope this work to interventions that preserve the designer's original layout: recommendations keep text in its existing position over its existing background, rather than proposing changes such as repositioning text or restructuring the layout. Layout-level interventions remain a valuable but distinct direction, which we revisit in Section~\ref{sec:73}}.
\end{enumerate}

\section{Implementation}
We developed a prototype canvas editor for creating visual content inspired by and mirroring aspects of Adobe Express (Figure~\ref{fig:teaser}). Adobe Express caters to professionals with limited design expertise through its template-driven approach and simplified toolset,\footnote{Template-driven design tools have gained adoption due to their ability to democratize design by providing pre-made, customizable layouts and elements  https://www.pcmag.com/news/adobe-unveils-creative-cloud-express-template-driven-design-tools} similar to other popular design platforms like Canva,\footnote{http://canva.com} enabling users to create polished visual content without extensive design training.

The \system{} canvas editor allows users to design visually rich documents while receiving just-in-time alerts to accessibility issues and actionable recommendations to improve accessibility. A dedicated side panel was added to display these alerts and recommendations, enabling us to study their impact on the authoring process. 

\subsection{User Interface}
Figure~\ref{fig:teaser} shows an overview of the user interface. The user interface of our system was developed using React for the front-end and a Flask server for the back-end. At the core of the canvas editing functionality, we used Konva.js,\footnote{https://konvajs.org/} a library for rendering interactive graphics on the web. Our implementation enables users to design visually rich content with intuitive drag-and-drop features on a canvas, allowing for easy addition and customization of shape and text elements. 

The left panel of the application provided essential editing functionalities, such as modifying font style, size, color, and outline (Figure~\ref{fig:teaser}a). The right panel focused on accessibility-related features (Figure~\ref{fig:teaser}b-c). This panel displayed just-in-time alerts for components with identified color contrast issues. The panel included an educational explanation of the alerts. For each element that received an alert, several recommendations were provided to help users resolve the issues.

\subsection{Algorithm Design \& Implementation}

\subsubsection{Color Contrast}

The system provided the user a warning any time the color contrast ratio ($R$) between the foreground text color and background colors fell below a minimum level. Unlike traditional contrast checkers that assume solid backgrounds, our approach addresses the challenge of complex backgrounds by sampling all background colors within the text rendering area. Given the complex backgrounds in our study, we chose the minimum ratio when computed for all background colors within 1 pixel of the text rendering. This approach ensures accurate contrast assessment even with non-uniform backgrounds, addressing limitations identified in prior work where popular contrast checker solutions misrepresent non-solid backgrounds and yield inaccurate pass/fail WCAG conditions~\cite{almeida2020analysis}.

We follow the WCAG 2 guidelines for computing the color contrast ratio as follows, where ($Y_1$) is the luminance of the lighter color and ($Y_2$) is the luminance of the darker color.

%\todo{technically,$Y_1$ is the lighter color, and $Y_2$ is the darker. Is the assumption that foreground is always lighter than the background? We should say something about it.}. 

\[ R = \dfrac{  Y_1 + 0.05 }{ Y_2 + 0.05}  \]

%%% Re ordered figure to match text
\begin{figure*}
    \centering
    \includegraphics[width=0.95\textwidth, 
    alt={Five variations of text styling for contrast improvement, all showing the phrase 'The quick brown fox jumps over the lazy dog' on a patterned background with curved brown and blue shapes. Two variations do not meet the target contrast ratio: white text (R=1.0), black text (R=2.0). The remaining variations meet contrast guidelines: outlined text (R=4.9), text with outline opacity (R=4.7), text with partial outline (R=10.2), background opacity (R=4.7), and solid background color (R=5.0). The contrast ratio (R) is indicated for each variation.}]{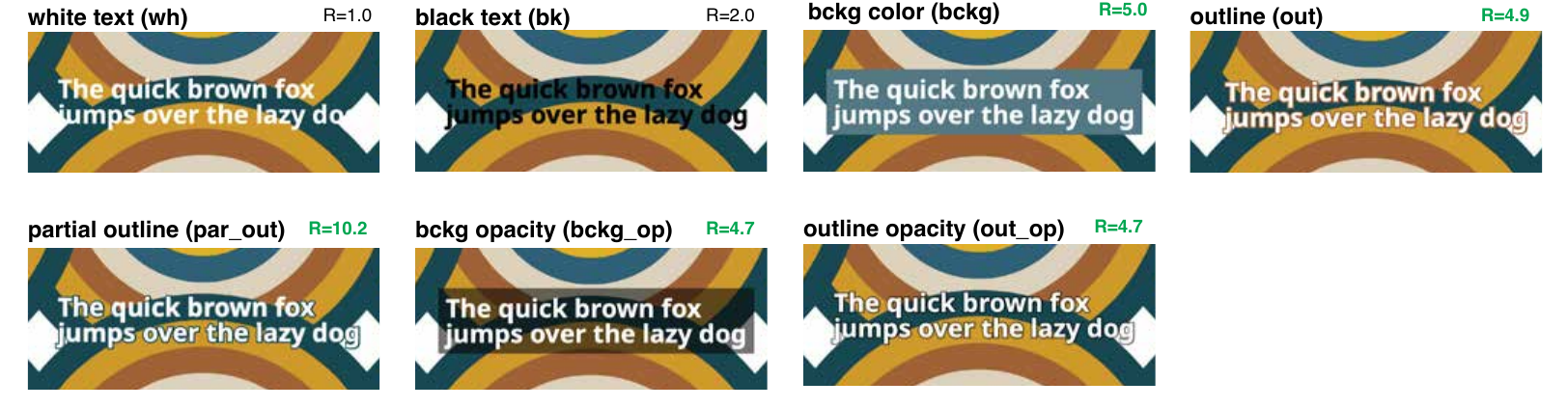}
    \caption{\changedcamera{White text (wh) and black text (bk) are shown for reference -- these do not meet the WCAG target contrast ratio of $4.5$. \system{} provides 5 different types of color recommendations to address a low contrast issue -- across text color, outline and opacity.}}
    \label{fig:recommendation_samples}
    \Description{Five variations of text styling for contrast improvement, all showing the phrase 'The quick brown fox jumps over the lazy dog' on a patterned background with curved brown and blue shapes. Two variations do not meet the target contrast ratio: white text (R=1.0), black text (R=2.0). The remaining variations meet contrast guidelines: outlined text (R=4.9), text with outline opacity (R=4.7), text with partial outline (R=10.2), background opacity (R=4.7), and solid background color (R=5.0). The contrast ratio (R) is indicated for each variation.}
\end{figure*}

\subsubsection{Color Recommendations}
\label{sec:color_recs}

When text with low contrast was found, the system provides the user with 6 different types of \changed{design} recommendations that meet the contrast requirements for WCAG text color contrast, as shown in Figure \ref{fig:recommendation_samples}. For each recommendation, up to 6 different colors could be recommended to the user within the prototype, as shown in Figure \ref{fig:teaser}. 
%% Justification of recommendations (R3)
\changedcamera{These recommendation types were informed by a review of how designers commonly address low-contrast text in practice. Two of the authors examined designer-made templates available within a production design tool (Adobe Express) to catalog strategies used beyond the default of switching to black or white text. This review surfaced recurring strategies including outlines, background color plates, opacity effects, and partial outlines. We note that this was a formative, exploratory review conducted by the authors rather than a formal expert elicitation study or a systematic, quantified analysis of strategy prevalence; it was intended to ground our recommendation set in existing design practice rather than provide a comprehensive taxonomy.}

\changed{A strength of the current technical method is that it generalizes to any design, allowing the user to place text on any complex background at any desired location in the document while still ensuring sufficient color contrast. This allows users to keep their desired template, background, or layout while choosing an accessible design aesthetic that meets their needs. This is critical in many common workflows where users often start with licensed or branded assets or templates in design tools such as Canva or Adobe Express.}

\begin{enumerate}

\item  \textbf{Text Color}:
% We generalize the approach from Adobe Color,\footnote{https://color.adobe.com/create/color-contrast-analyzer}
% to suggest text color changes from the same color family that meet a given contrast ratio. 
\changedcamera{We generalize the approach from Adobe Color,\footnote{https://color.adobe.com/create/color-contrast-analyzer} to suggest text color changes from the same color family\footnote{\changedcamera{We define color family as the perceptually related set of colors sharing a common hue/chroma region.}} that meet a given contrast ratio.}
Given a color of a pixel from a background $RGB_B$ and the foreground text color $RGB_T$, this approach recommends a new text color from the same color family that satisfies a contrast ratio to all adjacent background pixels. To do so, the text and background colors are first converted to the CIE $xyY$ color space, $xyY_B$ and $xyY_T$ respectively. %, where $Y$ is the relative luminance from above.  
Next, we solve for the minimum value(s) of $\bar{Y}_T$ that provides the required contrast value to all background pixels. %For efficiency purposes we can 

%\[ Y_2  = \dfrac{  Y_1 + 0.05 }{ R}   -  0.05 \] 

In some cases, the new $xy\bar{Y}_T$ value may be outside the available RGB gamut. If so, we find the closest RGB color by iteratively changing $xy$ in the direction of the white point. Note that in cases with complex backgrounds with varying luminance, there may not be a single text color that satisfies the contrast against all luminance values. In this case, the system does not recommend a color. 

\item  \textbf{Solid Background Color or Color Plate}:
Given a text color and the colors from a background, this approach recommends colors for a solid background below the text that provides enough contrast. Two types of colors are chosen to maintain the aesthetics of the design. In one type, we start from the color closest to the average from the background pixels adjacent to the text, and then use the approach above to find a variation with sufficient contrast with respect to the text. For the second type, we cluster the colors in the design and suggest the most common colors that have sufficient contrast from the text. For this study, only a background rectangle was use, though the approach generalizes to any background shape. 

\item  \textbf{Solid Outline Color}: This recommendation follows the color strategy for background color suggestions from the previous approach (Solid Color Background), but instead applies it as a 2 pixel outline. 

\item  \textbf{Partial Outline}: This recommendation also follows the color selection and outline strategy of the previous approach (Solid Outline Color). However, instead of applying the outline around the full text, it is only applied to the areas of the background that have low contrast. The intuition for this recommendation is that, in some cases, only a small portion of the background may be responsible for having the text be marked as low contrast.

\item  \textbf{Background Opacity}: In some cases, users may want to preserve the background content, which is obscured when using a solid background or outline. In this approach, we instead use an alpha mask for either a white (for dark text) or black (for light text) rectangular background to reach the desired text contrast. To find the ideal alpha value, we identify the lowest contrast color value from the background surrounding the text and solve for the alpha value that results in the desired text contrast ratio. 

\item  \textbf{Outline Opacity}: This color recommendation applies an outline with opacity, using the algorithms for an optimal alpha value from the previous approach. This provides a better blending with the background in comparison to a solid color outline.

\end{enumerate}
\section{Study 1: Recommendation Preferences}
\label{sec:study_1_survey}

We conducted a study to evaluate how different color recommendations affect both readability and aesthetic appeal in various design contexts. While all our color recommendations (discussed in Section~\ref{sec:color_recs}) achieve WCAG contrast requirements, we sought to understand which approaches users prefer and how these preferences vary across different background types (e.g., images, gradients, textures). This evaluation would inform how to prioritize and present recommendations in \system{}.

% How do human perceptual ratings of color contrast align with the predicted outputs of our algorithm?
% Are there preferences between the various proposed strategies to increase contrast (i.e., text color, outline, background, opacity, partial background)

\subsection{Methodology}
A within-subjects, repeated-measures design was used to present each participant with various design templates containing text under several conditions. The study was deployed online via a Qualtrics survey and designed for a duration of 25 minutes.

\subsubsection{Conditions}
Participants were exposed to multiple conditions to evaluate text readability and visual appeal. A control condition consisted of black/white maximal contrast with black or white font colors. Additionally, several interventions were implemented to improve contrast requirements; these included the six color recommendations outlined in Section~\ref{sec:color_recs}: 

\begin{enumerate}
    \item Changing the text color (txt)
    \item Adding an outline (out)
    \item Adding a solid background color (bckg)
    \item Adding a background opacity (bckg\_op)
    \item Adding a partial outline (par\_out)
    \item Adding a partial opacity outline (par\_op)
\end{enumerate}

For each intervention, the target contrast aimed to exceed the WCAG minimum of 4.5:1. Note that the (txt) intervention was not shown for image and gradient, as no text color could be found to meet this contrast ratio.  Examples of each intervention are shown in Figure~\ref{fig:recommendation_samples}. 

\begin{figure*}
    \centering
    \includegraphics[width=0.85\textwidth, 
    alt={Three categories of complex design backgrounds used to test ColorA11Y recommendations. The categories are labeled as 'texture', 'image', and 'gradient'. The texture category shows two examples: a repeating circular pattern in brown and blue tones, and a stylized group of human figures in various colors against a striped background. The image category displays two photographs: one of a silhouetted figure in a misty setting, and another showing a group portrait of four people. The gradient category features two examples of smooth color transitions: one with pastel rainbow colors and another with bold red, green, and blue blends.}]{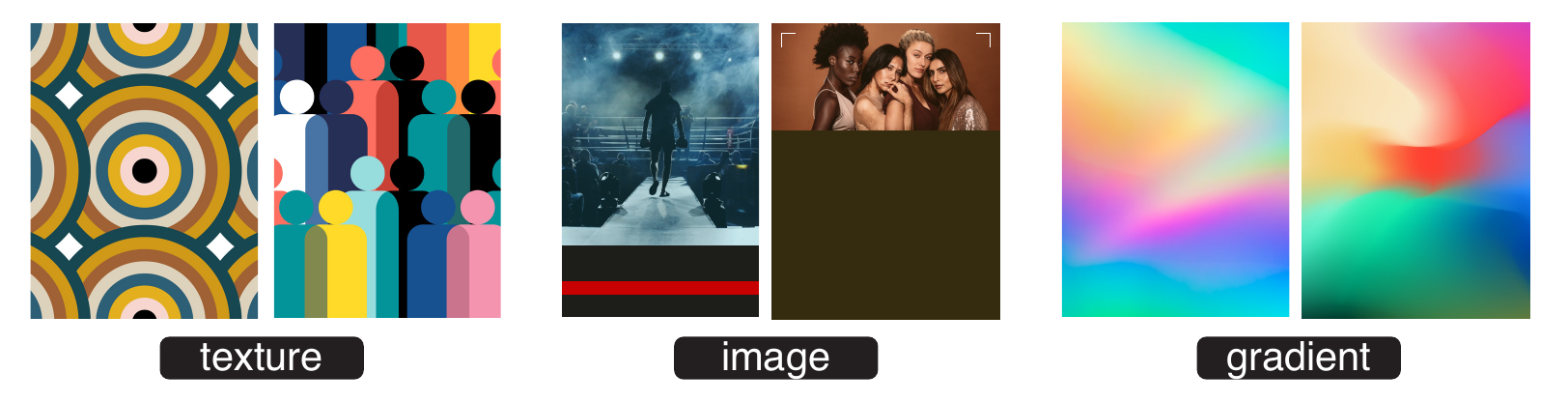}
    \caption{Design backgrounds used to obtain user preferences on different recommendations provided by \system.}
    \label{fig:survey_backgrounds}
    \Description{Three categories of complex design backgrounds used to test ColorA11Y recommendations. The categories are labeled as 'texture', 'image', and 'gradient'. The texture category shows two examples: a repeating circular pattern in brown and blue tones, and a stylized group of human figures in various colors against a striped background. The image category displays two photographs: one of a silhouetted figure in a misty setting, and another showing a group portrait of four people. The gradient category features two examples of smooth color transitions: one with pastel rainbow colors and another with bold red, green, and blue blends.}
\end{figure*}

\subsubsection{Stimuli}
Participants evaluated design backgrounds that varied in type, including: textures, images, and gradients. In total, 6 different backgrounds were used (Figure~\ref{fig:survey_backgrounds}). Each design was presented in combination with the various interventions for enhancing text contrast. The type of font used was also varied between serif (Times) and sans serif (Noto Sans), each presented at a size of 20 points. These font types were chosen based on prior work demonstrating their effectiveness for readability~\cite{wallace2022towards}; as such, the impact of more diverse font types was left out of scope for this study. The stimuli spanned a 3x3 factorial design background with additional between-subjects variations in font type. 

\subsubsection{Procedure}
Participants were presented with text (\textit{“The quick brown fox jumps over the lazy dog”}) on a series of design backgrounds, each incorporating different text interventions to enhance contrast. For each combination, participants rated the designs on two metrics along a 1-5 Likert scale: perceived text \textbf{readability} and \textbf{visual appeal}. They were also asked to provide free-form comments regarding what aspects of the design were working well and what could be improved. After evaluating each combination individually, participants ranked all the interventions. They were asked to identify their top two preferred options and their two least preferred options. Each participant saw one version of each design background type, paired with either a serif or sans serif font, and the order of interventions was randomized to reduce order effects.

 %% rev section
\subsection{\changed{Recruitment \& Participants}}
% Participants were recruited via Prolific\footnote{http://prolific.com} and screened to have specific characteristics relevant to the study. Specifically, they were required to create visual content regularly and use design tools such as Adobe Express or Canva. This ensured that participants had practical experience with visual design tasks, making their evaluations more informed and relevant to the context of the study. In total, 40 participants participated, and each design × intervention combination was rated by at least 20 different individuals. Participant compensation rate was 18 USD per hour.
\changed{Participants were recruited via Prolific
and screened to have specific characteristics relevant to the study. The main inclusion criteria for the study required participants to regularly engage in the design and creation of visual content and frequently use design tools such as Adobe Express or Canva. Recruitment was conducted in two stages to identify qualified participants without revealing the study's focus on accessibility. We first launched a general screening survey titled ``Work Preferences'' that included multiple-choice questions about work habits and tool usage (Appendix~\ref{sec:study_1_screening}). This survey was designed to take approximately 1 minute and did not reveal the study's purpose. A total of 750 Prolific users completed this initial screener. Based on our inclusion criteria, 107 Prolific users who met the requirements were invited to participate in the main study. Of these, 40 participants completed the study. This two-stage recruitment process ensured that participants had practical experience with visual design tasks, making their evaluations more informed and relevant to the context of the study. 70\% of the participants used design tools at least once a week. There was a range of expertise levels with 57.5\% rating themselves as having intermediate design proficiency and an additional 22.5\% having advanced or Expert level design proficiency. The most common tools used by the users were Canva (90\%), PowerPoint (57.5\%), and Photoshop (55\%). More detailed demographics for the participants can be found in Appendix A.2 and Figure ~\ref{fig:demographics}. Each design × intervention combination was rated by at least 20 different individuals. Participant compensation rate was 18 USD per hour.
} %% end rev section

\begin{figure*}
    \centering
    \includegraphics[width=0.85\textwidth, 
    alt={Likert scale ratings for text readability across different ColorA11Y recommendations and background types (gradient, image, and texture). The visualization is split into two parts for each background type: left side shows stacked bar charts displaying percentage distribution of responses from 'Strongly disagree' to 'Strongly agree', while right side shows dot plots with means and 95\% confidence intervals. The data compares eight text treatments: white text (wh), text (txt), partial outline (par_out), partial opacity (par_op), outline (out), black text (bk), background opacity (bckg_op), and background (bckg). The study included 40 participants (n=40) rating the statement 'The text is easy to read' for each condition. For gradient, txt has the largest group of Strongly Agree. For image, bckg. And for texture, bckg as well. Appendix Table 2 summarizes the visualization data in table format.}]{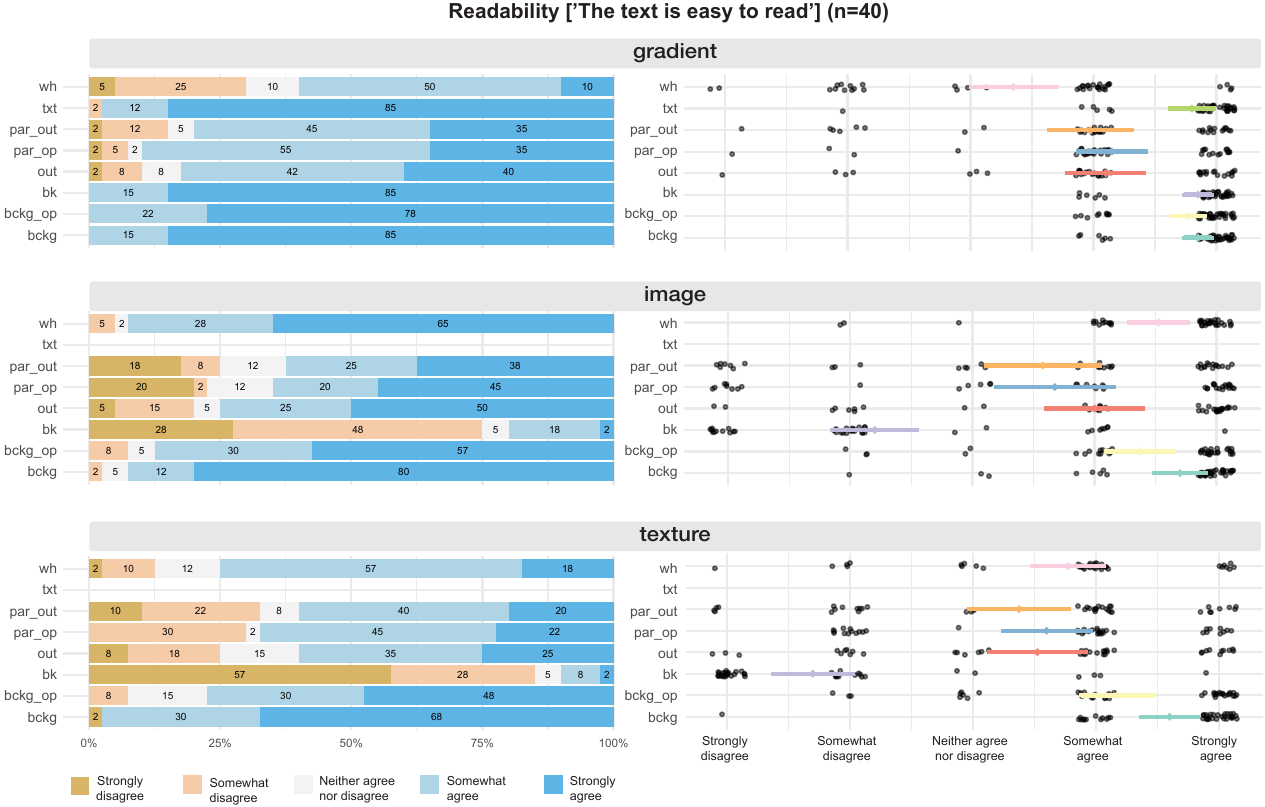}
    \caption{Likert ratings for readability of different recommendations for each design background (i.e., gradient, image, and texture). Left shows distribution percentage of responses, and right shows mean and 95\% CI.}
    \label{fig:likert_ratings_read}
    \Description{Likert scale ratings for text readability across different ColorA11Y recommendations and background types (gradient, image, and texture). The visualization is split into two parts for each background type: left side shows stacked bar charts displaying percentage distribution of responses from 'Strongly disagree' to 'Strongly agree', while right side shows dot plots with means and 95\% confidence intervals. The data compares eight text treatments: white text (wh), text (txt), partial outline (par_out), partial opacity (par_op), outline (out), black text (bk), background opacity (bckg_op), and background (bckg). The study included 40 participants (n=40) rating the statement 'The text is easy to read' for each condition. For gradient, txt has the largest group of Strongly Agree. For image, bckg. And for texture, bckg as well. Appendix Table 2 summarizes the visualization data in table format.}
\end{figure*}

\begin{figure*}
    \centering
    \includegraphics[width=0.85\textwidth, 
    alt={Likert scale ratings for visual appeal of the design across different ColorA11Y recommendations and background types (gradient, image, and texture). The visualization is split into two parts for each background type: left side shows stacked bar charts displaying percentage distribution of responses from 'Strongly disagree' to 'Strongly agree', while right side shows dot plots with means and 95\% confidence intervals. The data compares eight text treatments: white text (wh), text (txt), partial outline (par_out), partial opacity (par_op), outline (out), black text (bk), background opacity (bckg_op), and background (bckg). The study included 40 participants (n=40) rating the statement 'The design is visually appealing.’ for each condition. Most of the mean values overlap in CI. Only bk, stands out as lower for image and texture. Appendix Table 2 summarizes the visualization data in table format.}]{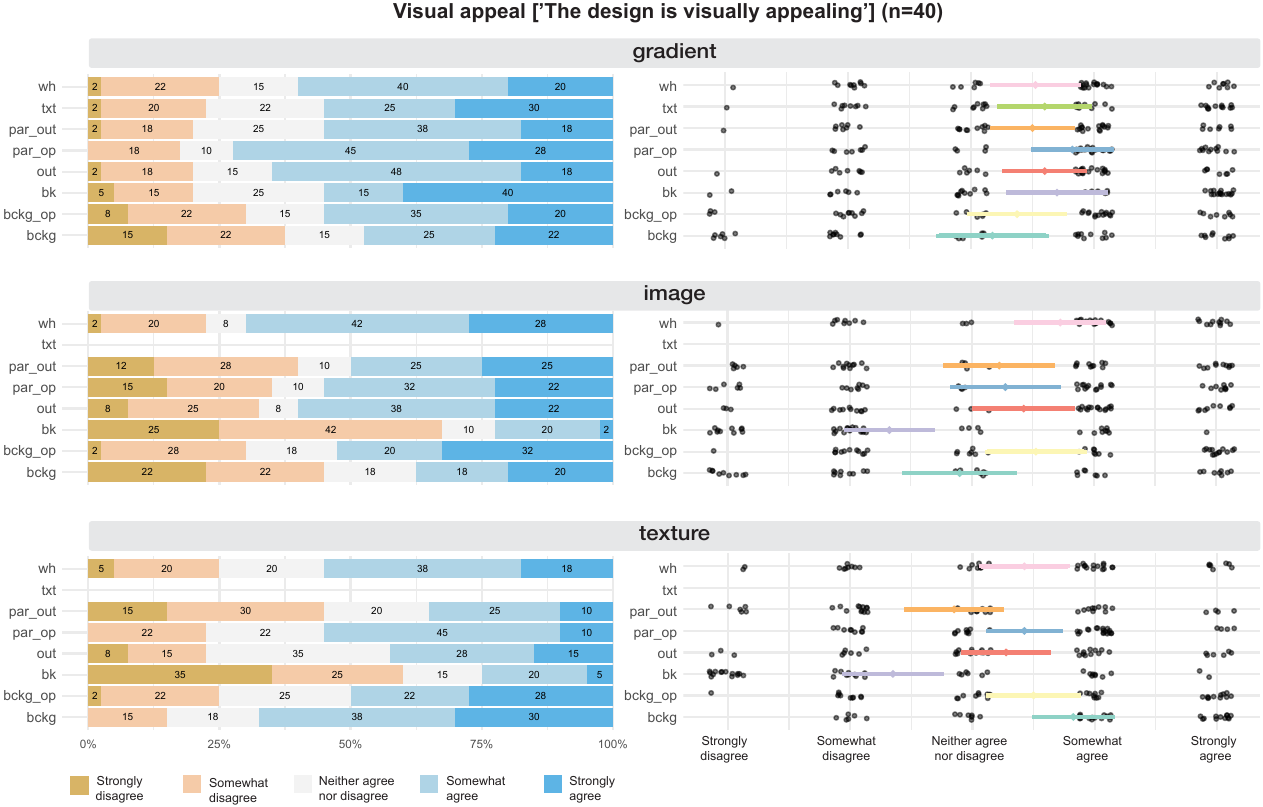}
    \caption{Likert ratings for visual appeal of different recommendations for each design background (i.e., gradient, image, and texture). Left shows distribution percentage of responses and right shows mean and 95\% CI.}
    \label{fig:likert_ratings_aes}
    \Description{Likert scale ratings for visual appeal of the design across different ColorA11Y recommendations and background types (gradient, image, and texture). The visualization is split into two parts for each background type: left side shows stacked bar charts displaying percentage distribution of responses from 'Strongly disagree' to 'Strongly agree', while right side shows dot plots with means and 95\% confidence intervals. The data compares eight text treatments: white text (wh), text (txt), partial outline (par_out), partial opacity (par_op), outline (out), black text (bk), background opacity (bckg_op), and background (bckg). The study included 40 participants (n=40) rating the statement 'The design is visually appealing.’ for each condition. Most of the mean values overlap in CI. Only bk, stands out as lower for image and texture. Appendix Table 2 summarizes the visualization data in table format.}
\end{figure*}

\begin{figure*}
    \centering
    \includegraphics[width=0.80\textwidth, 
    alt={Bar chart comparing user preferences for different text interventions across three background types (gradient, image, and texture). For each background type, users voted for their Best and Worst preferred recommendations, shown as percentages. The chart lists eight interventions: text (txt), background (bckg), background opacity (bckg_op), black text (bk), outline (out), partial opacity (par_op), partial outline (par_out), and white text (wh). Each row shows two percentage values and corresponding bars - blue for Best votes and brown for Worst votes. For gradient backgrounds the best two were text color (19\%) and background opacity (18\%) and the worst two were white text (24\%) and background color (20\%). For image backgrounds the best two were white text (29\%) and background opacity (29\%) and the worst two were black text (54\%) and background color (22\%). For texture backgrounds the best two were background color (53\%) and background opacity (20\%) and the worst two were black text (55\%) and partial outline (18\%). It is worth nothing that each of the color recommendation except for white text, black text, and partial outline has more then five percent of users mark it as a best choice for all three backgrounds, while only background was chosen by more then 50\% of users as their best choice for textured backgrounds.}]{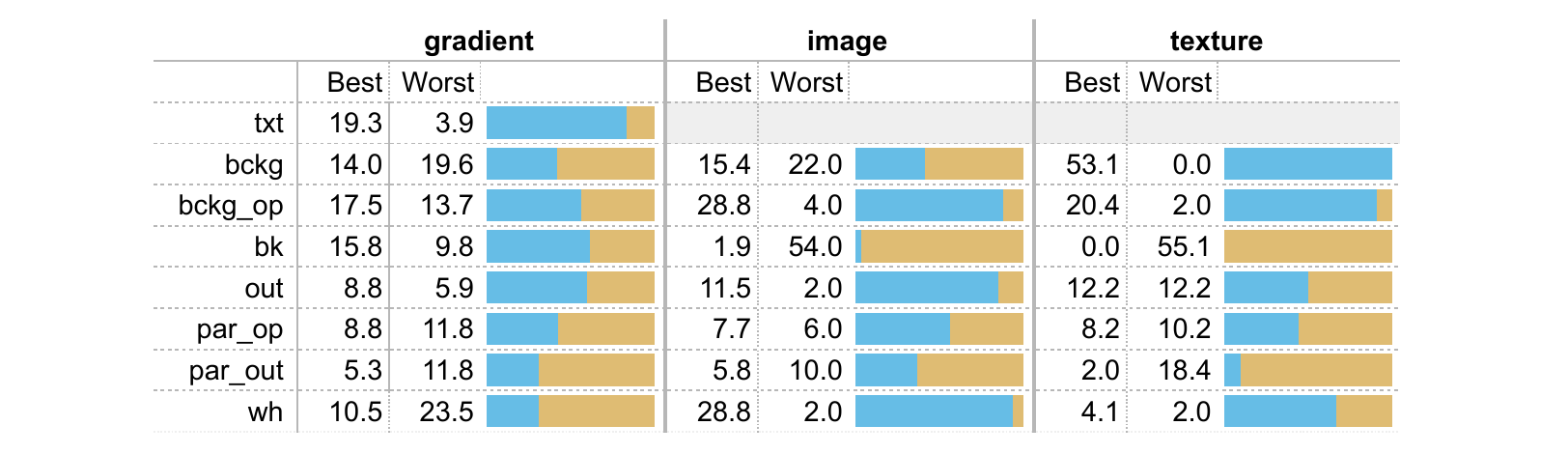}
    \caption{Percentage (\%) of votes for Best and Worst intervention. For each background type, users picked their most preferred recommendation (Best) and least preferred recommendation (Worst).}
    % In balancing Best and Worst, users opted for recommendations that made the text readable, matched the style of the design, and did not take away attention from the background (especially for images).
    \label{fig:ranking_table}
    \Description{Bar chart comparing user preferences for different text interventions across three background types (gradient, image, and texture). For each background type, users voted for their Best and Worst preferred recommendations, shown as percentages. The chart lists eight interventions: text (txt), background (bckg), background opacity (bckg_op), black text (bk), outline (out), partial opacity (par_op), partial outline (par_out), and white text (wh). Each row shows two percentage values and corresponding bars - blue for Best votes and brown for Worst votes. For gradient backgrounds the best two were text color (19\%) and background opacity (18\%) and the worst two were white text (24\%) and background color (20\%). For image backgrounds the best two were white text (29\%) and background opacity (29\%) and the worst two were black text (54\%) and background color (22\%). For texture backgrounds the best two were background color (53\%) and background opacity (20\%) and the worst two were black text (55\%) and partial outline (18\%). It is worth nothing that each of the color recommendation except for white text, black text, and partial outline has more then five percent of users mark it as a best choice for all three backgrounds, while only background was chosen by more then 50\% of users as their best choice for textured backgrounds.}
\end{figure*}

\subsection{Data Analysis \& Model Selection}

We analyzed two key measures: readability and aesthetic appeal, using cumulative link mixed models (CLMM) with participant ID as a random effect to account for repeated measures. We tested several nested models to identify the best fit for our data. The final selected model included:
\begin{itemize}
    \item A fixed effect for intervention type (7 levels: background color, background opacity, outline, partial opacity, partial outline, black text, white text)
    \item A fixed effect for design background type (3 levels: gradient, image, texture)
    \item The interaction between intervention type and design background
    \item A random effect for participant ID to account for repeated measures
\end{itemize}

This model provided the best fit for both readability ($AIC = 1770.1$, $BIC = 1888.4$, $df=25$) and aesthetics ($AIC = 2417.7$, $BIC = 2536.1$). While we tested models including font type (serif vs. sans-serif) as an additional predictor, likelihood ratio tests did not show significant improvements to model fit (readability: $p = 0.147$; aesthetics: $p = 0.204$), suggesting that font choice had minimal impact on the effectiveness of accessibility interventions and user preference.

To understand the specific differences between interventions within each design context, we conducted post-hoc pairwise comparisons using estimated marginal means (\textit{emmeans} in R) with Tukey's adjustment for multiple comparisons. Effect sizes were calculated as odds ratios from the model estimates.

\subsection{Results}

\subsubsection{Likert Ratings}

The effectiveness of readability and visual appeal of different accessible interventions varied depending on the design background type (gradient, image, or texture). Figure~\ref{fig:likert_ratings_read} shows aggregated Likert ratings for readability and Figure~\ref{fig:likert_ratings_aes} shows ratings for visual appeal. Appendix Table~\ref{table:pairwise_read} and \ref{table:pairwise_aes} shows pairwise contrast p-values.

For image backgrounds, solid background color achieved the highest readability ($M = 4.81$, $SE = 0.08$) but received relatively low visual appeal ratings ($M = 2.96$, $SE = 0.23$). Background opacity provided a good compromise, maintaining good readability ($M=4.47$, $SE=0.14$) and somewhat higher visual appeal ($M=3.61$, $SE=0.20$). 
% Given the potentially \textit{busy} nature of images, interventions that cover the image improve readability but are less visually appealing.

For gradient backgrounds, multiple interventions received significantly higher readability ratings: solid background ($M=4.89$, $SE=0.053$), background opacity ($M=4.82$, $SE=0.073$), and black text color ($M=4.89$, $SE=0.05$). Ratings for visual appeal were consistent across interventions with no significant pairwise differences. These suggest that gradient backgrounds are more robust to different intervention choices along both measures. 

Unlike other design contexts, texture backgrounds showed alignment between readability and visual appeal ratings. Solid background color (readability: $M=4.67$, $SE=0.11$; visual appeal: $M=3.83$, $SE=0.19$) and background opacity (readability: $M=4.27$, $SE=0.15$; visual appeal: $M=3.54$, $SE=0.199$) were rated as clearly preferred for both metrics. Pairwise contrasts showed no significant difference between background and background opacity for both metrics (Figure~\ref{fig:likert_ratings_read} and \ref{fig:likert_ratings_aes}). In particular, background color was significantly better in readability compared to all other interventions (Figure~\ref{fig:likert_ratings_read} and \ref{fig:likert_ratings_aes}). 

The performance of standard text color change (black or white) varied markedly across interventions, challenging their use as default solutions. For readability, black was rated highly for gradient ($M=4.89$, $SE=0.05$) and poorly for images ($M=2.23$, $SE=0.21$) and texture ($M=1.62$, $SE=0.16$). Visual appeal followed similar pattern, high ratings for gradient ($M=3.80$, $SE=0.191$) and lower ratings for images ($M=2.35$, $SE=0.19$) and textures ($M=2.27$, $SE=0.19$). White text showed similar contextual dependency (Figure~\ref{fig:likert_ratings_read} and \ref{fig:likert_ratings_aes}).

\subsubsection{Overall Rankings}

Figure \ref{fig:ranking_table} shows the percentage of participants that chose their best two or worst two preferences for each color recommendation. For gradient backgrounds the best two were text color (19\%) and background opacity (18\%) and the worst two were white text (24\%) and background color (20\%). For image backgrounds the best two were white text (29\%) and background opacity (29\%) and the worst two were black text (54\%) and background color (22\%). For texture backgrounds the best two were background color (53\%) and background opacity (20\%) and the worst two were black text (55\%) and partial outline (18\%). Its worth nothing that each of the color recommendation except for white text, black text, and partial outline has more then five percent of users mark it as a best choice for all three backgrounds, while only background was chosen by more then 50\% of users at their best choice for textured backgrounds.

\subsubsection{Qualitative Comments}

Qualitatively, participants described their preferences through a lens that considered both readability and aesthetic quality, with these factors often in tension with each other. 
%This tension was particularly evident in how participants evaluated different interventions across background types.
When assessing readability, participants valued clear contrast and legibility, yet consistently rejected solutions that achieved this at the expense of design integrity across different background types. 

For image backgrounds, while a solid background color provided maximum contrast and was described as most readable, they were also criticized for their \textit{"lack of blend"}, being \textit{"invasive"} and \textit{"taking away from the background"}. Background opacity was described as a preferred intervention that better achieved this balance, with participants praising how they made \textit{"text clearer and easy to read while still showcasing some of the image behind it."} This qualitative feedback aligned with the quantitative ratings where background opacity maintained good readability while achieving higher visual appeal than solid backgrounds. On the other hand, for textured backgrounds, participants favored solid backgrounds that provided clear separation from busy patterns. While for gradient backgrounds, participants favored text color modifications which felt \textit{"in harmony"} and \textit{"works with the gradient"}.

Across design backgrounds, partial outline emerged as a particularly controversial intervention. Participants described it as \textit{"awful"}, \textit{"atrocious"} and \textit{"weird"}; though were described more positive when paired with sans serif fonts. While plain white text often offered high contrast against darker backgrounds, some participants expressed concern in its fit with the overall design, suggesting that it felt \textit{"basic"}, \textit{"amateurish"} and \textit{"jarring"}.

\subsection{Takeaways}

\changed{While WCAG provides general contrast guidelines, little research exists on user preferences for different interventions when multiple compliant options are available. This study aimed to understand user preferences for different accessible interventions across complex backgrounds with the following findings: }

\begin{enumerate}
    \item \textbf{Context matters significantly for accessibility interventions.} The effectiveness of different accessible interventions varies substantially by design background type, challenging the assumption that any WCAG-compliant solution is equally acceptable. 
    % Based on readability and visual appeal measures as well as user rankings, the effectiveness of different accessible interventions varies substantially by design background type. 
    Design integration stood out as a key factor, with participants strongly preferring interventions that felt purposeful rather than merely functional. 
    \item \textbf{No single intervention works best across all contexts}. While current accessibility guidelines focus on technical compliance, they provide limited design guidance for maintaining visual appeal across different background types. A solid background color or background with some opacity is often a good choice for offering accessibility improvements across design contexts. Standard approaches of using black or white text perform inconsistently, challenging their use as the only default solutions. Based on consistent negative feedback, we removed partial outlines from \system's recommendations in its current implementation.
    % Partial outline received consistent negative feedback from participants and is best avoided in its current implementation.
    \item \textbf{Multiple compliant options better serve designers' diverse aesthetic needs}. While readability ratings often had a more clear winner, visual appeal ratings often varied. This suggests that accessibility guidelines could benefit from context-specific recommendations that prioritize a suite of compliant options rather than prescribing single solutions. Providing designers with multiple accessible choices accommodates different aesthetic preferences while maintaining compliance, supporting both accessibility goals and creative freedom.
    % This suggests the value of providing users with multiple accessibility options that accommodate different aesthetic preferences for designers.
\end{enumerate}
\section{Study 2: Workflow Experience Evaluation}
\label{sec:study_2_experience}

We conducted a second study to understand the impact of our tool on the users' experience while authoring design content. Specifically, we had the following objectives:
\begin{enumerate}
    \item Understand users' current challenges in authoring accessible designs.
    \item Assess whether \system{} helps users identify and correct accessibility issues more effectively compared to baseline color contrast checker tools.
\end{enumerate}

The evaluation compared the user's experience during a design task using our tool and a typical color contrast checker tool in Adobe Express. We hypothesized our tool can 
can improve the authoring experience of accessible design by reducing the effort needed for remediation decisions.
% reduce users' effort in making a design accessible by minimizing design decisions related to remediating accessibility issues. 
Additionally, our tool can increases users' confidence in ensuring their design meets accessibility guidelines, particularly for users with limited knowledge of these guidelines.

\begin{table*}[]
\centering
\caption{Participant biographies and self-reported design and accessibility practices}
\label{table:participants}
\Description{Participant demographics and their experience with design and accessibility. The table contains information for 8 participants (P1-P8) with columns showing their professional title, age, frequency of content creation, design proficiency level, and familiarity with accessibility concepts. The participants range in age from 22 to 54 years old and hold various professional roles including IT Consultant, Graphic Designer, Producer, Life Coach, Teachers, Owner, and President. Four participants create content daily, while the other four do so several times a week. Their design proficiency levels vary from Novice to Expert, with most participants at the Intermediate or Advanced level. Regarding accessibility familiarity, two participants are 'Very familiar' with accessibility concepts, five are 'Somewhat familiar', and one is 'Slightly familiar'.}
\begin{tabular}{llrlll}
\rowcolor[HTML]{C0C0C0} 
{\color[HTML]{101010} \textbf{ID}} &
  {\color[HTML]{101010} \textbf{\changedcamera{Occupation}}} &
  \multicolumn{1}{l}{\cellcolor[HTML]{C0C0C0}{\color[HTML]{101010} \textbf{Age}}} &
  {\color[HTML]{101010} \textbf{\begin{tabular}[c]{@{}l@{}}Frequency\\ creating content\end{tabular}}} &
  {\color[HTML]{101010} \textbf{\begin{tabular}[c]{@{}l@{}}Design \\ Proficiency\end{tabular}}} &
  {\color[HTML]{101010} \textbf{\begin{tabular}[c]{@{}l@{}}Accessibility\\ Familiarity\end{tabular}}} \\
\rowcolor[HTML]{FFFFFF} 
P1 &
  {\color[HTML]{101010} IT Consultant} &
  {\color[HTML]{101010} 54} &
  {\color[HTML]{101010} Daily} &
  {\color[HTML]{101010} Advanced} &
  {\color[HTML]{101010} Very familiar} \\
\rowcolor[HTML]{F3F3F3} 
P2 &
  \cellcolor[HTML]{FFFFFF}{\color[HTML]{101010} Graphic Designer} &
  \cellcolor[HTML]{FFFFFF}{\color[HTML]{101010} 22} &
  {\color[HTML]{101010} Daily} &
  {\color[HTML]{101010} Expert} &
  {\color[HTML]{101010} Very familiar} \\
\rowcolor[HTML]{FFFFFF} 
P3 &
  {\color[HTML]{101010} Producer} &
  {\color[HTML]{101010} 27} &
  Daily &
  {\color[HTML]{101010} Novice} &
  Somewhat familiar \\
\rowcolor[HTML]{F3F3F3} 
P4 &
  \cellcolor[HTML]{FFFFFF}{\color[HTML]{101010} Life Coach} &
  \cellcolor[HTML]{FFFFFF}{\color[HTML]{101010} 33} &
  Daily &
  Intermmediate &
  Slightly familiar \\
\rowcolor[HTML]{FFFFFF} 
P5 &
  Teacher &
  {\color[HTML]{101010} 25} &
  Several times a week &
  Intermmediate &
  Somewhat familiar \\
\rowcolor[HTML]{F3F3F3} 
P6 &
  Owner &
  38 &
  Several times a week &
  Intermmediate &
  Somewhat familiar \\
\rowcolor[HTML]{FFFFFF} 
P7 &
  Teacher &
  35 &
  Several times a week &
  {\color[HTML]{101010} Novice} &
  Somewhat familiar \\
\rowcolor[HTML]{F3F3F3} 
P8 &
  President &
  37 &
  Several times a week &
  {\color[HTML]{101010} Advanced} &
  Somewhat familiar
\end{tabular}
\end{table*}

\subsection{Participants}

Participants were recruited via UserInterviews.com\footnote{https://www.userinterviews.com/} and compensated 50 USD for their time. Each session had an average duration of 72 minutes (SD = 13.5). Table~\ref{table:participants} provides a summary of participant demographics and self-reported expertise in design and familiarity with accessibility guidelines. 

Our participants represented a diverse range of experience levels and professional contexts, primarily consisting of content creators who manage visual content for their own small businesses (6 out of 8 participants), and professional designers who work with client requirements (2/8). Self-reported experience in creating visual content varied from Novice (2/8) to Intermediate (3), Advanced (2), and Expert (1). Most participants created content for social media and marketing purposes and often used tools such as Adobe Express to meet their needs.

\subsection{Procedure}

The interviews were conducted online using a videoconferencing tool and recorded and transcribed for further analysis. The primary data collection included qualitative feedback from participants during the design tasks, followed by a semi-structured interview regarding their experiences and preferences between the two tools. We also collected quantitative measures such as user satisfaction with their final result, confidence in meeting accessibility standards, and measures of usability assessed using the After-Scenario Questionnaire (ASQ)~\cite{lewis1991psychometric}. \changed{We additionally collected the task and time performance for the design task. Please note that the time performance was impacted by the think-aloud procedure, although participants were asked to do this for both conditions, and we also ask about their subjective efficiency in the questionnaire. The task performance measured how many of the text element's users created met WCAG guidelines color contrast. }

For the design tasks, participants completed two open-ended design tasks in a counterbalanced manner across two conditions:
\begin{enumerate}
    \item Baseline: Participants used Adobe Express and a typical color contrast checker add-on which required participants to probe for specific foreground/background colors (Appendix Figure~\ref{fig:baseline}). After users probe specific colors using a color picker, the tool provided a color contrast value and a pass/fail assessment based on WCAG guidelines.
    \item \system: Participants used our tool, which provided just-in-time alerts to accessibility issues and recommended color choices to improve accessibility while they were authoring content (Figure~\ref{fig:teaser}).
\end{enumerate}

In both tasks, participants were provided with text and a design background (either an image or texture) and were instructed to create a draft of their design (Appendix Figure~\ref{fig:experience_templates}). They were asked to make the design aesthetically pleasing while also ensuring that it met accessibility guidelines. This process involved setting the layout and making font and color decisions to achieve both visual appeal and accessibility.

\begin{figure*}
    \centering
    \includegraphics[width=0.85\textwidth, 
    alt={Ratings compared between ColorA11Y and baseline conditions across five metrics (ASQ - satisfaction with ease of task completion, ASQ - satisfaction with tool support received, ASQ - satisfaction with task completion time, confidence in meeting color contrast accessibility guidelines, and design satisfaction). The visualization presents data in two formats: stacked bar charts showing percentage distributions on the left, and mean values with 95\% confidence intervals on the right. Each metric uses a five-point Likert scale from 'Strongly disagree' to 'Strongly agree'. The results generally favor ColorA11Y over baseline, with notably higher positive ratings for confidence in meeting accessibility guidelines and satisfaction with task completion time.}]{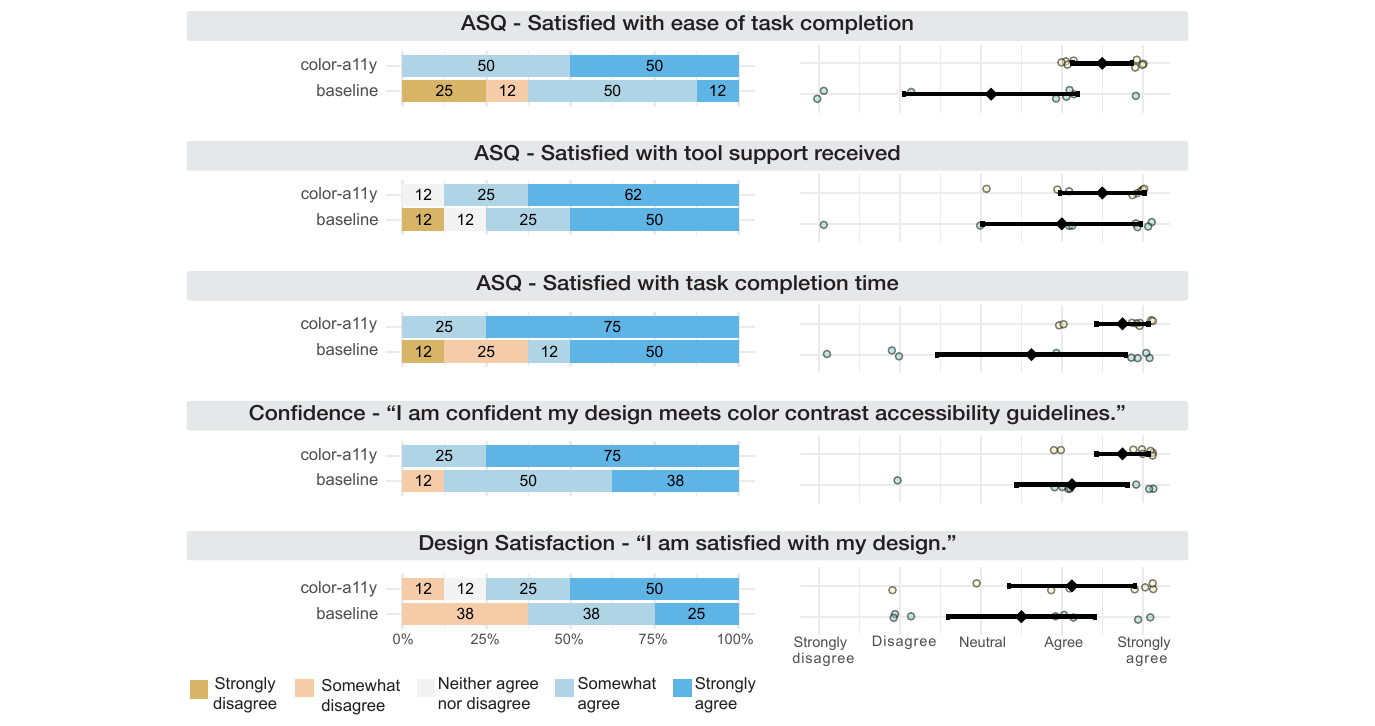}
    \caption{Likert ratings for ASQ Usability Questionnaire and post-task completion metrics. Left shows distribution percentage of responses and right shows mean and 95\% CI.}
    \label{fig:likert_asq}
    \Description{Ratings compared between ColorA11Y and baseline conditions across five metrics (ASQ - satisfaction with ease of task completion, ASQ - satisfaction with tool support received, ASQ - satisfaction with task completion time, confidence in meeting color contrast accessibility guidelines, and design satisfaction). The visualization presents data in two formats: stacked bar charts showing percentage distributions on the left, and mean values with 95\% confidence intervals on the right. Each metric uses a five-point Likert scale from 'Strongly disagree' to 'Strongly agree'. The results generally favor ColorA11Y over baseline, with notably higher positive ratings for confidence in meeting accessibility guidelines and satisfaction with task completion time.}
\end{figure*}

\begin{table}[h!]

\centering

%\makebox[\textwidth]{%
%\colorbox{revisionblue}{
%\parbox{0.95\textwidth}{%

\caption{\changed{Comparison of the task and time performance for the Study 2 users for Color-A11y and the baseline. Time performance measure the time taken by the user to complete the design and task performance measures the percentage of elements created by users that met WCAG requirements for color contrast.}}
\Description{Time and task performance for Color-A11y in comparison to the baseline system. The times are reported in hours:minutes:seconds and the task performance is reported as a percentage of text elements that meet WCAG color contrast guidelines}

\label{table:task_performance}

\begin{center}
{\changed{\begin{tabular}{l|cc|cc}
\hline
 & \multicolumn{2}{c|}{Color-A11y Performance} & \multicolumn{2}{c}{Baseline Performance} \\
Participant & Time & Task & Time & Task \\
\hline
P1 & 0:20:14 & 100\% & 0:20:16 & 100\% \\
P2 & 0:06:55 & 100\% & 0:07:00 & 100\% \\
P3 & 0:17:51 & 100\% & 0:12:10 & 67\% \\
P4 & 0:08:40 & 100\% & 0:06:18 & 100\% \\
P5 & 0:12:50 & 100\% & 0:15:33 & 100\% \\
P6 & 0:04:06 & 100\% & 0:04:01 & 0\% \\
P7 & 0:03:50 & 100\% & 0:07:31 & 67\% \\
P8 & 0:10:13 & 100\% & 0:17:26 & 33\% \\
\hline
Average & 0:10:35 & 100\% & 0:11:17 & 71\% \\
\hline
\end{tabular}}}
\end{center}

%}%
%}
%}%

\Description{  }
\end{table}

\subsection{Findings}

\subsubsection{Current Accessibility Practices}
% Most participants reported limited familiarity with accessibility guidelines, ranging from slight (P4) to moderate understanding (P5, P6, P7, P8), with only two participants (P1, P2) expressing confidence in applying accessibility best practices.
Participants described strategies they employed to enhance content accessibility of their creations. All eight participants described some form of readability considerations. Techniques employed included using bold formatting, increasing font sizes, adjusting font or background colors to improve contrast, and adding text shadows or outlines (P1-P8). Some participants also considered cognitive accessibility through careful placement of elements and management of text density (P1, P3, P5, P6, P7). 
 
Participants described several situations where they encountered tensions between aesthetic goals and accessibility guidelines. Those familiar with accessibility guidelines noted that following specific requirements, such as minimum font sizes, often compromised their desired visual design (P3, P5, P7). As P3 described it: 
\textit{``Sometimes what's accessible visually, is not compatible with the overall design and art of it… like if guidelines say, to make the font size a certain size, but aesthetically it looks better smaller. That's an issue.''}
These tensions manifested in various contexts. P2 described situations where they had to reconcile client brand guidelines with accessibility requirements. Others faced technical constraints when working with specific design assets (P4, P5, P6). For instance, P4, who creates content for a spiritual coaching business, described the challenge of maintaining text legibility over colorful crystal imagery: 
\textit{``There's been some images that I've wanted to use, that it was just almost impossible to get the text to be legible over that image.''}

\subsubsection{Seamless Integration Into Authoring Workflow}
\label{sec:findings_seamless_workflow}

Our evaluation demonstrated that \system{} successfully integrated accessibility considerations into users' natural workflow. In the After-Scenario Questionnaire, participants \changedcamera{subjectively} rated \system{} highly for ease of use (4.5/5), time efficiency (4.8/5), and support (4.5/5)---significantly higher than the baseline tool's ratings of 3.1, 3.6, and 4.0 respectively (Figure~\ref{fig:likert_asq}). 
{\changed{For task performance shown in Table \ref{table:task_performance}, all designs and elements created with ColorA11Y met WCAG color-contrast guidelines, compared to only 50\% of designs and 71\% of elements in the baseline condition, despite similar completion times (10:35 with ColorA11Y vs. 11:17 with the baseline). As noted earlier the time performance may have been impacted by the think-aloud approach, but note that users subjectively found themselves more efficient (4.8/5 vs 3.6/5) with  ColorA11Y.}} 
Qualitatively, participants highlighted three key benefits: a streamlined workflow, just-in-time feedback, and non-disruptive design guidance.

\textbf{Streamlined Workflow.} \system{} reduced accessibility effort through a more streamlined workflow by eliminating the tedious process of manually checking individual elements (P1, P2, P3, P4, P5, P7, P8). In the baseline condition, participants had to repeatedly select elements, probe specific colors, and make adjustments without clear guidance. P5 described this challenge: 
\textit{``I wish that when I were to click the font or like click whatever box I'm on if it would just populate there... instead of me going in and like doing the eyedropper tool to pick the color every single time.''}
\system's automated detection removed these manual steps, with P1 noting: 
\textit{``It's more streamlined. It helps you walk through your project flow quicker. So that's a value, anything that saves time in the design process.''}
Many users either incorrectly used the baseline tool by probing wrong foreground and background colors, checking only parts of their design, or were uncertain how to evaluate complex image backgrounds. As P4 noted,
\textit{``it would make things so much easier and quicker for me rather than going through a bunch of different colors to figure it out.''}.

\textbf{Timely feedback.} The system's just-in-time feedback also proved particularly effective in supporting the user's workflow. By providing accessibility information exactly when and where users needed it during their design process, the tool eliminated the need for separate evaluation steps. As P8 highlighted: 
\textit{``I like that a lot how it was just instantly updating. You didn't have to stop what you were doing.''}
The integration of feedback into their current design process meant users could immediately see and address accessibility issues without disrupting their creative flow. While the baseline tool provided useful information, participants noted it was likely to be forgotten since it required active and conscious engagement (P1, P3, P5, P8). Users expressed reluctance to invest time in manual checking, with P8 explaining: 
\textit{``unless it was like automatic or some way to make it faster, I don't think I'd want to waste a bunch of time on checking to make sure.''}

\textbf{Design guidance.} The majority of participants perceived the alerts as helpful nudges rather than interruptions to their work, noting how they visibly led to improvements in their designs (P2, P3, P4, P7, P8). As P7 explained: 
\textit{``I mean it is disruptive, but in a good way... I'm okay with it being disruptive as long as it's helping me fix an issue.''}
This sentiment was echoed by multiple participants who found the tool's integration natural and beneficial to their workflow. P3 directly compared the two approaches: 
\textit{``this [\system] was less disruptive and easier to use to my workflow, more integrated. It felt more cohesive for sure.''}

\subsubsection{Scaffolding Accessibility Knowledge Through Actionable Recommendations}
\label{sec:findings_scaffold_recommendations}

\system{} effectively supported users in both identifying and resolving accessibility issues through its recommendation system. This scaffolding addressed two key challenges faced by our participants: limited knowledge of accessibility guidelines and varying levels of design expertise. The tool provided support for these at two levels: helping users recognize accessibility issues in their designs and guiding them toward appropriate solutions.

First, the tool helped build awareness of accessibility considerations in users’ design work. Six participants (P2, P4, P5, P6, P7, P8) specifically noted the educational value of learning to recognize accessibility issues in their work. P2 reflected: 
\textit{``I really like it. I would love to have the alerts actually because, like, even when I was going through and using the tools I expected the contrast level to be good. And it wasn't.''}

Second, the tool's actionable recommendations helped bridge the gap between identifying problems and implementing solutions. This was especially valuable for participants less confident in design work (P1, P5, P7). As P7, who has limited accessibility familiarity, explained: 
\textit{``Just so much easier. It kind of does a much better job explaining to me what the issues are, and just so much better just kind of providing those recommendations and helping me conduct the actual tasks to fix the issue.''}
This approach contrasted sharply with the baseline tool, which participants described as leaving them to \textit{``figure it out on your own"''} (P7). By providing actionable solutions, the tool helped participants move from understanding accessibility problems to confidently implementing fixes in their designs.

\subsubsection{Balancing Creative Control with Accessibility Requirements}
\label{sec:findings_creative_control}

Our evaluation revealed both successes and tensions in balancing creative control with accessibility requirements. The recommendations served as effective creative starting points, helping participants move beyond basic solutions while maintaining accessibility (P2, P3, P7). P2 noted: 
\textit{``I do like how in [\system] it recommended colors cause I feel like my automatic ideas, just to go back to white or black.''} \system{} helped users see a range of interventions they could apply to address accessibility issues (e.g., text color modifications, adding backgrounds or outlines, modifying opacity levels, etc).
This scaffolded approach to creativity enabled participants to explore accessible designs while maintaining their creative vision, with P7 explaining it was \textit{``so much easier to kind of get a good idea of what's going to work and then from there I could still make it my own.''}

However, participants also identified areas where creative control felt constrained. Participants described the baseline tool as providing more freedom for color exploration through the manual color picker tool (P2, P4). In contrast, \system’s recommendations could be limited, and some participants wanted more influence to steer the color palette of recommendations (P1, P2, P4, P6). P2 expressed: \textit{``I do think having the option to pick and try my own colors outside of what's recommended is probably the most important thing cause I would like to have that creative freedom.''} Beyond color selection, participants desired additional creative controls for design elements such as background dimensions, corner rounding, and opacity adjustments (P3, P8) -- features which were not supported in \system.

\section{Discussion}

\changed{\system{} closes the gap designers face between creative aesthetic constraints and adherence to color accessibility guidelines through a system that provides just-in-time feedback and actionable recommendations for complex real-world documents. We discuss our findings in relation to prior work and implications for design tools and accessibility practice.}%
\subsection{\changed{Aesthetic preferences for color accessibility are context dependent}}

\changed{ Prior work has indicated the importance of color in visual design in terms of the behavioral and emotional responses that can be elicited from users~\cite{valdez1994effects, eiseman2000pantone,ling2002effect, cyr2010colour, bonnardel2011impact}. Yet, the variety of ways in which color can be used in design creates risk of color being used in inaccessible ways. }

%Therefore, it becomes important that designers have adequate resources that can balance the exploration of aesthetic color preferences and accessible colors in their work~\cite{tigwell2017ace}. 
\changed{A recent scoping review found that most color accessibility design tools developed through academic research do not support designers with guidance on how to fix color issues (e.g., often the tools will highlight low contrast elements or use color vision deficiency simulations to highlight where colors will be indistinguishable)~\cite{geddes2025designing}. Our first study (Section~\ref{sec:study_1_survey}) highlighted the importance of providing multiple accessibility recommendations as users' preferences for balancing readability and aesthetics varied significantly across different design contexts.  This was also echoed in our second Study (Section~\ref{sec:study_2_experience}), where designers indicated they appreciated the design guidance and recommendations to push beyond simple fixes, but still wanted to personalize the final design changes.}

\changed{Our results resonate with palette-level work such as~\cite{tigwell2017ace}, where designers valued exploration of aesthetic color preferences and accessible colors but were constrained by limited integration into workflows. \system{} extends this line of work by showing that design tools must not only detect insufficient contrast but also provide multiple viable, contextually grounded solutions within the context of the design. These findings also suggest that automated solutions must go beyond simple color pairings or single best color fix and adapt to the context of the design as well as the designer's style. Designers benefited from a multitude of design and color options (i.e. text, outline, background, opacity). Finally, our findings on design contexts suggest a strong need for a more streamlined process in providing recommendations, which reduces cognitive load for design exploration and app overhead~\cite{hegemann2024palette} by providing context-driven color choices during the designing process. 
}

%, given the issue that designers also usually have to access a multitude of external tools~\cite{grobelny2015, almeida2020analysis}

%Therefore, \system{}'s approach has addressed a need for design tools that offer active guidance to address accessibility issues.}} 

%{\changed{}}

\subsection{\changed{Towards Born Accessible Design}}

\subsubsection{\changed{Color Accessibility Should Be Part of the Authoring Process}}

\changed{While previous tools provide some color accessibility support, they often focus on specific tasks, such as selecting compliant color palettes or supporting the checking of the luminance contrast between two colors~\cite{Coolors, iWantHue, colormaker, ColorBrewer2}}. \system{} demonstrates how color accessibility can be meaningfully integrated into the creative design process through a \emph{born-accessible} approach~\cite{lazar2023framework, lazar2026bornaccessible}{\changed{, as opposed to approaching accessible design with post hoc evaluations~\cite{vienot1999digital,reinecke2016enabling,padure2020comparing}. Addressing accessibility later in the design process is costly~\cite{horton2024techbrief,wentz2011retrofitting}. Furthermore, prior work has highlighted the frequent occurrence of inaccessible color use online~\cite{kuzma2010accessibility,patra2014quantitative,pillai2022websites,webaim2021the}, suggesting that current design tools and guidelines are not providing adequate accessibility support before digital products are released.}} \system{} showcases the potential of providing proactive, context-aware guidance throughout the authoring workflow that improves accessibility, rather than treating it as a final checkpoint.  {\changed{Built-in features for color contrast have significant design implications --- compared to optional third-party plugins --- as suggested by the recent Figma feature release on proactive color contrast checking,\footnote{\url{https://www.figma.com}. Figma announced the feature in March 2025.} however \system{} addresses the gap when it comes to suggestions for complex backgrounds.}}

\subsubsection{\changed{Can Documents be Made Born Accessible?}}%
\changed{For over a decade, disability rights activists have called for born-accessible design~\cite{wentz2011retrofitting}, and government policies are now starting to require it~\cite{adaTitle2,doitNDmaryland}, despite the details of how to do it not yet being in place~\cite{lazar2023framework, lazar2026bornaccessible}. Prior work has discussed the challenge where accessibility is viewed as disruptive to the creative process~\cite{andrew2025reframing}, signaling a need for creative accessibility design tools that can support meeting both accessibility and aesthetic design goals. Recent work has started to provide the details on methods and activities for born-accessible design~\cite{lazar2025opportunity}. However, tools are also necessary to support content creators in doing born-accessible design, and as far as we know, this is the first work that both creates a fully functional tool to support content creators in doing born-accessible design for color accessibility with complex background support, while documenting detailed feedback about the experience people have with using the tool. Our user studies establish the ease and time-savings with born-accessible authoring in real-time, such as avoiding manual color-checking of individual elements, and skipping tedious steps through proactive recommendations.} 
\changedcamera{Quantifying objective time savings is an important direction for future work.}

While our work focused on color contrast, we envision a born-accessible approach should consider all aspects of accessible content creation. The challenge lies in supporting multiple accessibility features (like reading order, alt text, and semantic structure) without overwhelming users or disrupting their workflow. Our findings suggest that the key lies in the timing and presentation of guidance; e.g., participants appreciated alerts that felt like helpful nudges rather than interruptions. This principle could guide the integration of additional accessibility features, perhaps using a progressive disclosure approach that introduces complexity as users become more comfortable with basic accessibility concepts. While more straightforward with color accessibility, this approach could be more difficult for other accessibility features or overwhelming when combined altogether. In a broader implementation, this difficulty can be addressed with a feature to mark elements as placeholders --- so as to suppress accessibility checks --- until they are replaced. 

Born-accessible design tools have the potential to transform how we create accessible digital content. As demonstrated through \system{}, integrating accessibility guidance directly into design workflows helps make accessibility an integral part of the creation process. Our work contributes to a growing understanding of how technical systems can support inclusive design practices while supporting designers' creative agency. By making accessibility guidance more contextual, timely, and aligned with creative workflows, born-accessible approaches could lead to a digital world where inclusive design becomes standard practice rather than an afterthought.

%Our two user studies revealed several benefits.  Second, our workflow study (Section~\ref{sec:study_2_experience}) demonstrated that just-in-time accessibility guidance can lead to effective outcomes, both providing a better authoring experience and resulting in more accessible content. {\changed{}} We found \system{} helped participants learn about and address accessibility issues without disrupting their creative workflow (Section~\ref{sec:findings_seamless_workflow}). This seamless integration is in contrast to traditional remediation approaches, where accessibility checks are altogether forgotten or often come too late in the process, leading to costly redesigns or compromised accessibility{\changed{~\cite{accessibilityFirst,wentz2011retrofitting}}}. Importantly, participants found the recommendations helped bridge the gap between identifying accessibility issues and implementing effective solutions, particularly valuable for those less familiar with accessibility guidelines and design practices (Section~\ref{sec:findings_scaffold_recommendations}).

\subsection{\changed{ Opportunities to Further Improve Color Accessibility}}
\label{sec:73}
While there was overall positive feedback using \system{}, our evaluation revealed further opportunities to enhance users' sense of creative control. 
Future iterations could allow users to refine or steer the recommendation algorithm through preferred color palettes or style guides, similar to how design systems operate. 
Additionally, providing preview and edit capabilities within the recommendation panel could help users explore variations while maintaining accessibility requirements, along with controls for design elements like background dimensions, corner rounding, and opacity adjustments that participants specifically requested (Section~\ref{sec:findings_creative_control}). These enhancements would further support the balance between accessibility compliance and creative freedom that participants valued.

Although the current implementation of \system{} relies on WCAG contrast ratio guidelines, we could expand its functionality in future work to integrate alternative color contrast guidelines. The choice to use WCAG was due to its global recognition as an accessibility standard and being referred to in different laws and design guidelines~\cite{Ara2024,apple_accessibility_vision,material_design_color_contrast,wcag_eaa_compliance_2024}. However, the way that we designed \system{} and what we identified was effective for users' design workflows means that we do not need to restrict \system{} to using WCAG's contrast ratio calculations. Some prior work has noted limitations with WCAG color contrast calculations and using it with complex backgrounds~\cite{almeida2020analysis,sharma_2024,oconnor2019orange} or certain color combinations, and alternative calculations have been proposed (e.g., the Accessible Perceptual Contrast Algorithm~\cite{myndex2022apcaeasyintro,moreno_designing_2024}). Alternative algorithms could be integrated without disrupting the workflow benefits we identified in our studies with \system{}. 
% {\color{red}{Future color systems should take a holistic account of these calculations especially if recommendations are produced by AI, as automatically generated content can exacerbate existing accessibility challenges and create new hurdles for visually impaired users, with significant downstream consequences of inaccessible AI-produced content~\cite{das_chi_2024}.}}

Moreover, participant feedback suggests opportunities for more sophisticated approaches that align with designers' natural workflows to extend our accessibility solution beyond color accessibility. Designers might consider text placement and visual hierarchy as part of their accessibility strategy~\cite{bajammal2021semantic,panchekha2018verifying,rello2012layout}. Some of our participants recognized this and described moving text to clearer areas of an image rather than just modifying colors. Future systems could incorporate these factors, recommending optimal text positions relative to complex backgrounds or suggesting adjustments to the visual hierarchy that maintain both accessibility and design intent. 
Additionally, recommendations could consider the relative importance of different text elements when suggesting accessibility improvements, preserving intended emphasis while ensuring readability. This multi-faceted approach could better align with how designers think about and solve accessibility challenges, moving beyond color modifications to support more holistic design decisions.
\changedcamera{These directions depart from the scope defined in Section~\ref{sec:design_goals} — moving beyond recommendations that preserve text's original position toward layout-level interventions that redesign where and how content is placed. We see this as a natural next step for later iterations of \system.}

\section{Limitations}

\subsection{Technical Limitations} 

In calculating color contrast, we \changed{adopt  the WCAG guidelines, which} make simplifying assumptions about color perception. The system uses a binary threshold for determining minimum contrast ratios and does not consider the relative proportion of background colors when making recommendations. This approach may not fully capture the nuanced ways humans perceive contrast in complex visual designs. Moreover, while our system provides accessible luminance-based color recommendations that meet WCAG color contrast requirements \changed{intended to aid users with low-vision and color deficiency}, it does not specifically optimize for color-blind users or other forms of color vision deficiency. Future work could incorporate more sophisticated color palette generation techniques that consider multiple forms of color vision impairment, building on prior work in this area~\cite{ribeiro2019recoloring}. Lastly, our solution does not address other important aspects of accessible design such as type scale, spacing, or semantic structure. A more comprehensive accessibility tool would need to consider these additional factors while maintaining the benefits of just-in-time feedback.

Another important consideration is the accessibility of the design tool itself. The focus of our work was to establish \system's fit within typical designer workflows, but we only recruited sighted designers to do this. 
Unfortunately, popular digital design tools and the way in which those tools require user interaction is often inaccessible~\cite{li2021accessibility,schaadhardt2021understanding}, which creates barriers to people with disabilities being designers, although researchers are looking to change this~\cite{piedade2025access,zhang2023a11yboard}. We could continue our work with \system{} by evaluating how well it fits into the design workflow of people with disabilities to identify further modifications we could make.

\subsection{Study Limitations} 

%Neither study controlled for participants' viewing environments, including screen settings, ambient lighting, or viewing distance. These factors can impact color perception and user preference~\cite{yu2016color}, potentially affecting both the ratings in Study 1 and experience in Study 2. 
While we recruited participants with varying levels of design expertise, there may be self-selection bias in our sample. Participants who chose to participate might have had a greater prior interest in or awareness of accessibility issues, potentially influencing their receptiveness to accessibility-focused tools compared to the broader designer population; particularly in the experience evaluation. We report participants' self-reported expertise in Table~\ref{table:participants}. \changed{Our studies focused on objective designer adherence to WCAG 2 color contrast guidelines. We did not specifically include individuals with low vision nor color vision deficiencies in the user population for the studies or evaluation, since it could be difficult to determine whether their subjective experiences arise from limitations in WCAG or limitations in ColorA11y. We will explore this further in future work.} 

In our experience evaluation, we had a limited and focused design task. This approach, while providing comparative data between the two tools evaluated, may not fully capture the complexities of real-world design workflows or long-term tool usage. In particular, the effectiveness and perceived utility of just-in-time alerts may differ in extended use scenarios or more complex design projects. {\changed{While our evaluation approach is common~\cite{remy2020evaluating}, a future study to evaluate \system{}'s long-term use would be a recommended next step}}. Additionally, the baseline condition provided access to full design tool capabilities, whereas our prototype had more limited editing features. This disparity in functionality, as well as participants' existing familiarity with the baseline tool, can also influence participants' workflow experiences and tool comparisons.
\section{Conclusion}

\changedcamera{We demonstrate that a \emph{born-accessible} approach to color accessibility can be compatible with current design workflows, leading to more accessible content without compromising the work}. By providing contextual guidance at the right moment, tools like \system{} can help bridge the gap between accessibility requirements and creative workflows.
Our findings reveal that designers value and benefit from just-in-time guidance that respects their creative vision while promoting accessible design practices. We hope this approach marks a shift towards born-accessible workflows, from treating accessibility as an afterthought to making it an integral part of the content creation workflow. As digital content continues to proliferate, tools that support born-accessible design could significantly impact content accessibility at scale, ultimately contributing to a more inclusive digital world. Future work building on these insights could extend beyond color contrast to address a wider range of accessibility considerations.

%TC:ignore
\begin{acks}
\changedcamera{We thank our participants for their time and insights. This work was supported in part by the University of Maryland Grand Challenges Program and Adobe Research.
This material is based upon work supported by the National Science Foundation under Grant No.~2333220 (awarded to Tigwell).}
\end{acks}
%TC:endignore

%TC:ignore
\bibliographystyle{ACM-Reference-Format}
\bibliography{references}
%TC:endignore

%%
%% Appendix
%TC:ignore
\appendix
\section{Study 1: Recommendation Preferences}

%\revsection{ %% Rev section
{\color{black}{
%{\color{blue}{
\subsection{Screening Survey}
\label{sec:study_1_screening}

\subsubsection*{Activity}
Which of the following tasks do you regularly perform as part of your work or personal endeavors?
\begin{itemize}
    \item Writing reports or essays
    \item Creating presentations or slideshows
    \item Analyzing data or spreadsheets
    \item Designing or creating visual content (e.g., flyers, social media content, posters, brochures, etc)
    \item Coding or programming
    \item Conducting research
    \item Managing social media accounts
    \item Editing photos or videos
    \item None of the above
\end{itemize}

\subsubsection*{Frequency}
How often do you create visual content?
\begin{itemize}
    \item Daily
    \item Several times a week
    \item Once a week
    \item A few times a month
    \item Once a month
    \item Every few months
    \item A few times a year
    \item Rarely or never
\end{itemize}

\subsubsection*{Tools}
To create visual content, which of the following software or tools have you used in the past month?
\begin{itemize}
    \item PowerPoint
    \item Canva
    \item Adobe Express
    \item Photoshop
    \item Illustrator
    \item Google Slides
    \item Keynote
    \item Prezi
    \item Sketch
    \item Inkscape
    \item Other. Please specify: \underline{\hspace{3cm}}
\end{itemize}

\subsubsection*{Proficiency}
How would you rate your proficiency in creating visual content?
\begin{itemize}
    \item Beginner --- I'm just starting out and learning the basics
    \item Novice --- I can create simple designs with guidance or templates
    \item Intermediate --- I can create decent designs independently for most of my needs
    \item Advanced --- I'm proficient with various design tools and can create complex visuals
    \item Expert --- I have extensive experience and can create professional-quality designs for any purpose
\end{itemize}

}}

%} %end revsection

% Option 2: Include as 2x2 grid of subfigures

\begin{figure*}[htbp]
    
    \centering
%\begin{tcolorbox}[colback=revisionblue, colframe=blue, boxrule=0pt,  enlarge left by=0mm, enlarge right by=0mm]
\begin{tcolorbox}[colback=white, colframe=white, boxrule=0pt,  enlarge left by=0mm, enlarge right by=0mm]
    
    \begin{subfigure}[b]{0.48\textwidth}
        \centering
        \includegraphics[width=\textwidth, 
            alt={Study 1 demographics consisting of four charts summarizing study user attributes as percentages of users: (a) - Frequency of tool usage: {A few times a month:15\%,Daily:20\%,Every few months:7.5\%,Once a month:7.5\%,Once a week:7.5\%,Several times a week:42.5\%)}}]{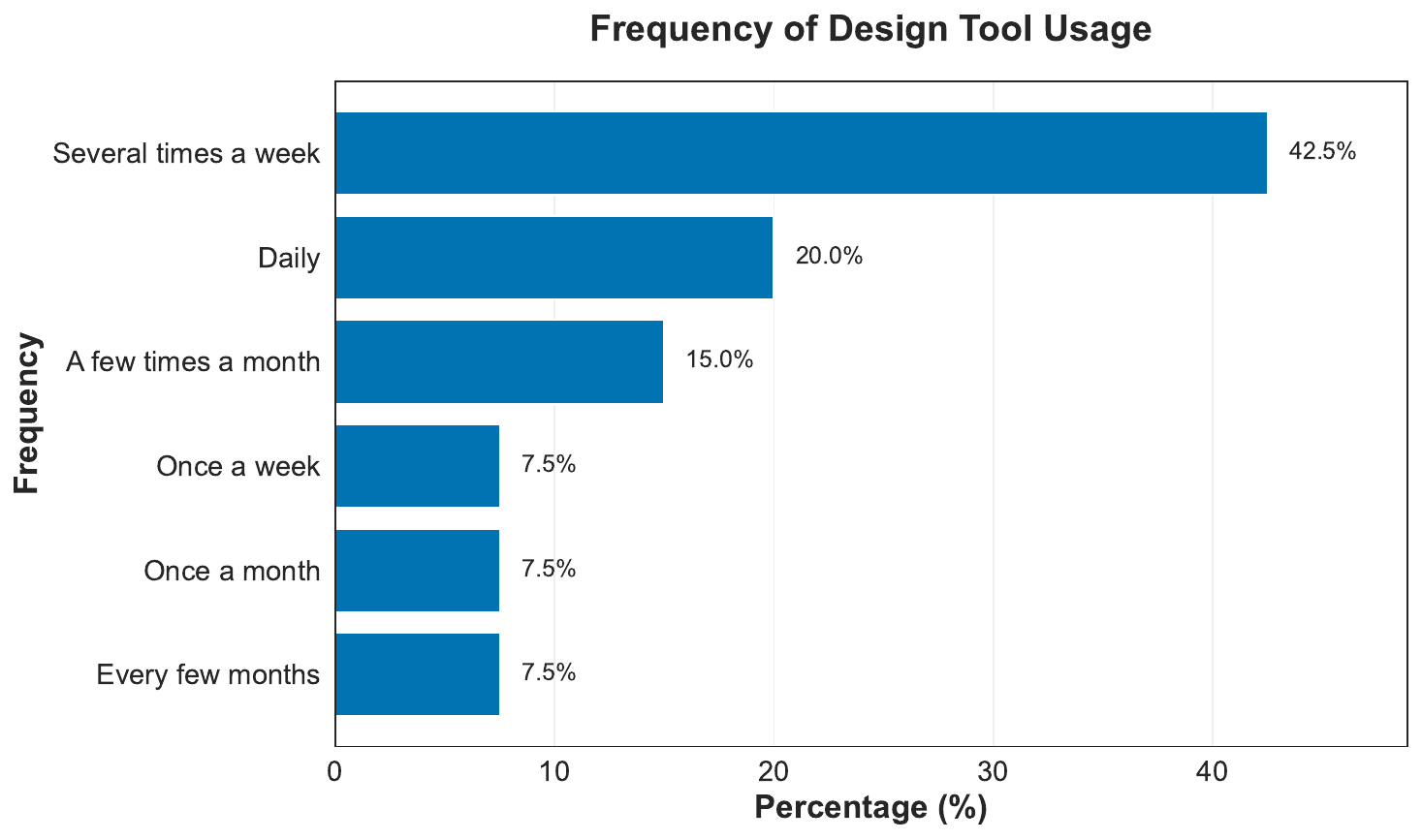}
        \caption{Usage frequency}
        \label{fig:freq_sub}
    \end{subfigure}
    \hfill
    \begin{subfigure}[b]{0.48\textwidth}
        \centering
        \includegraphics[width=\textwidth, 
            alt={(b) - Design Proficiency: {Advanced:15\%,Beginner:5\%,Expert:7.5\%,Intermediate:57.5\%,Novice:15\%}}]{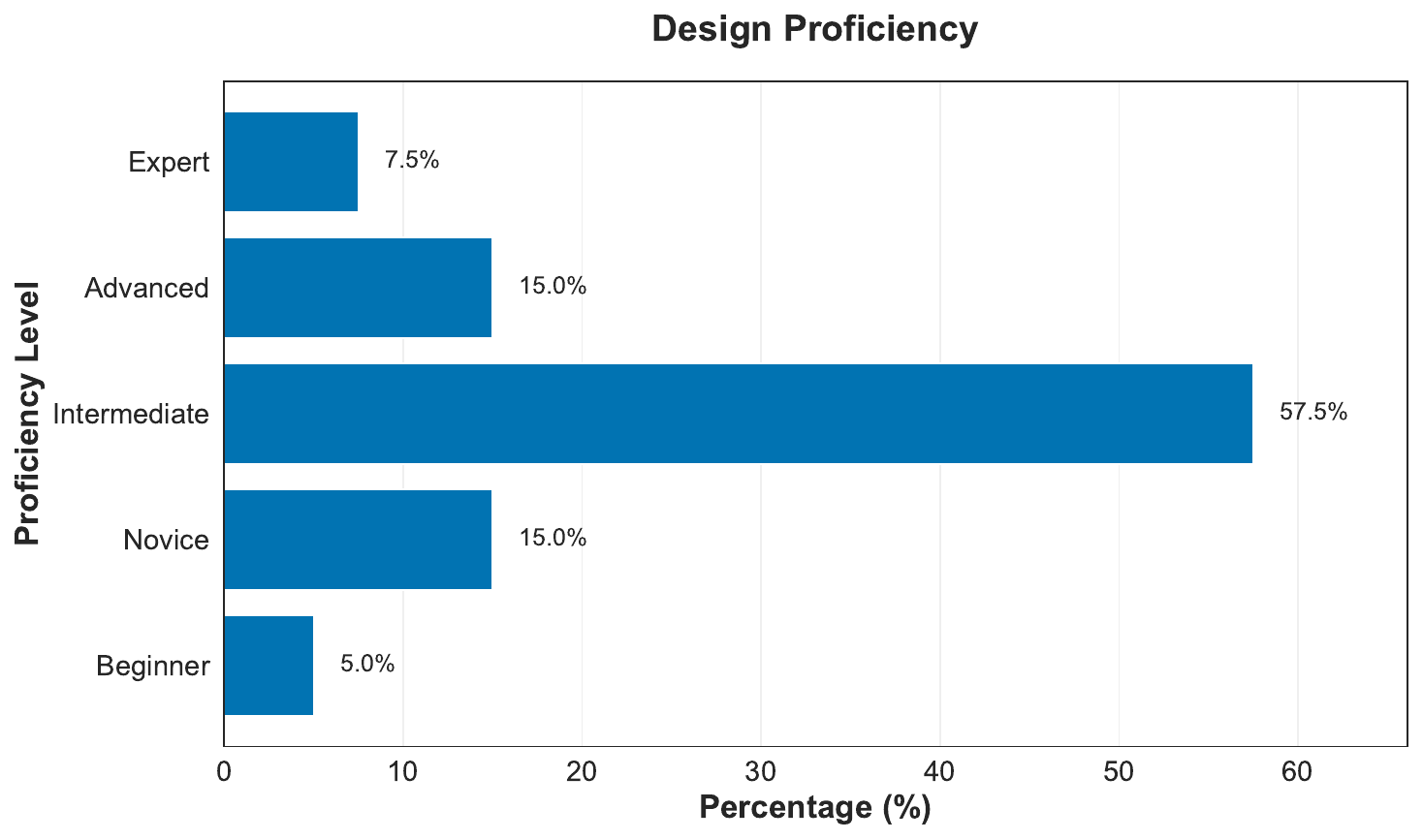}
        \caption{Proficiency levels}
        \label{fig:prof_sub}
    \end{subfigure}
    
    \vspace{1em}
    
    \begin{subfigure}[b]{0.48\textwidth}
        \centering
        \includegraphics[width=\textwidth, 
            alt={(c) - Design Tasks performed: {Designing or creating visual content:100\%, Editing photos or videos:77.5\%, Writing reports or essays:65\%, Conducting research:62.5\%, Creating presentations or slideshows:62.5\%, Analyzing data or spreadsheets:55\%, Managing social media accounts:50\%, Coding or programming:10\%, None of the above:2.5\%}}]{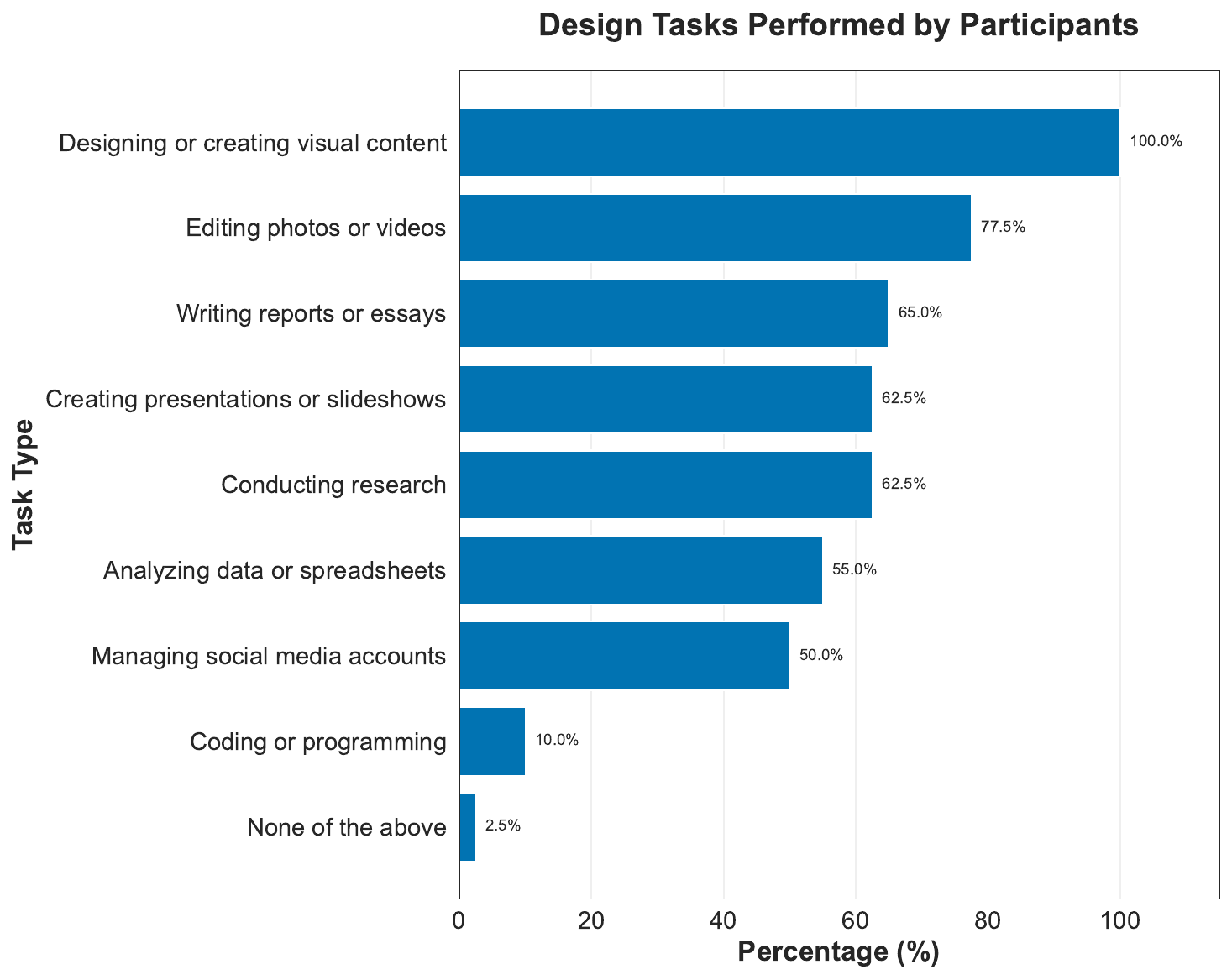}
        \caption{Common tasks}
        \label{fig:tasks_sub}
    \end{subfigure}
    \hfill
    \begin{subfigure}[b]{0.48\textwidth}
        \centering
        \includegraphics[width=\textwidth, 
            alt={(d) - Design tools used in the past month: {Canva:90\%, PowerPoint:57.5\%, Photoshop:55\%, Google Slides:45\%, Adobe Express:32.5\%, Illustrator:20\%, Keynote:17.5\%, Prezi:15\%, Inkscape:7.5\%, Other:7.5\%, Sketch:5\%}}]{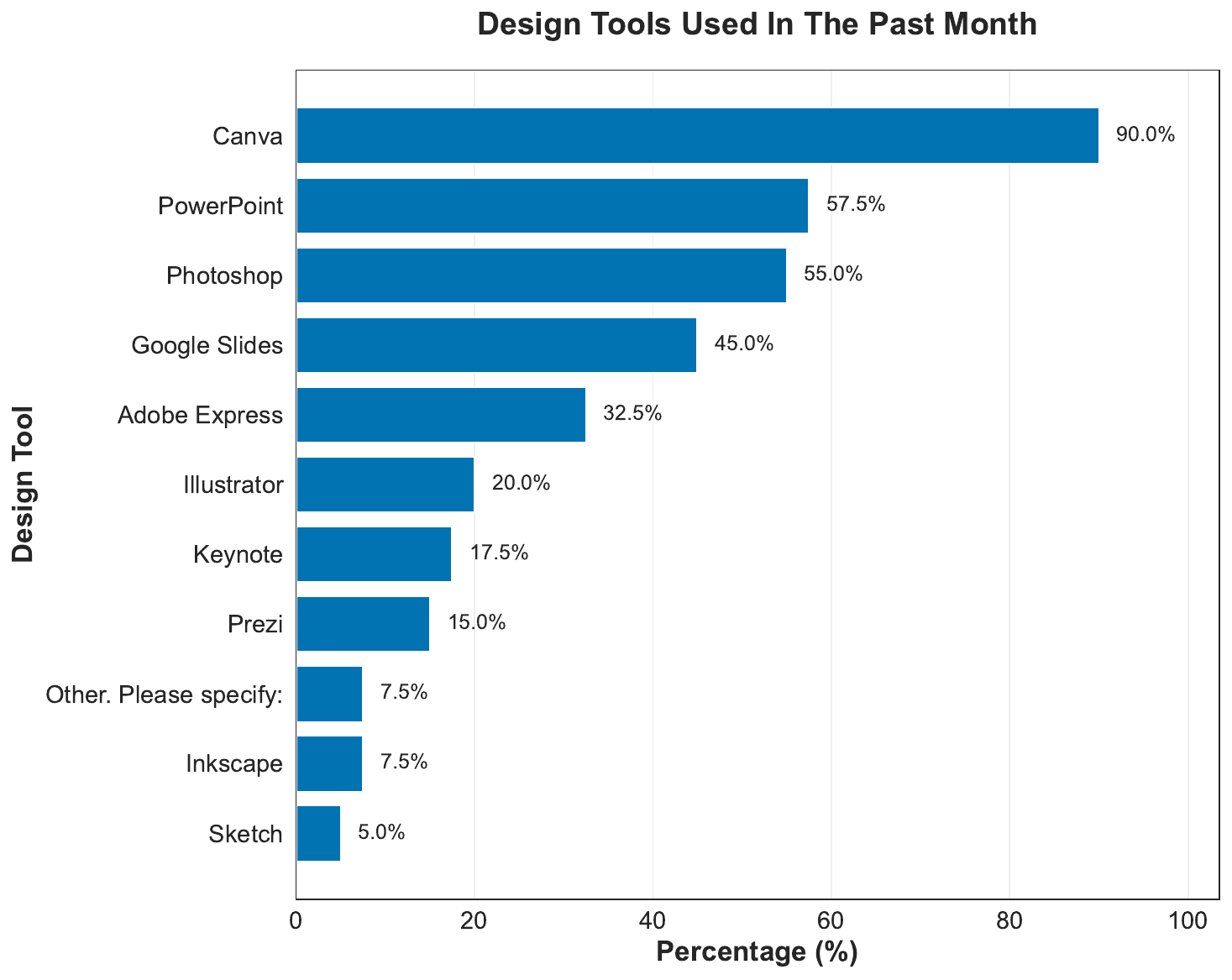}
        \caption{Tools used}
        \label{fig:tools_sub}
    \end{subfigure}
    
    \caption{Study 1 demographics (N=40).}
    \label{fig:demographics}
\end{tcolorbox}    
    
    \Description{Study 1 demographics consisting of four charts summarizing study user attributes as percentages of users: (a) - Frequency of tool usage: {A few times a month:15\%,Daily:20\%,Every few months:7.5\%,Once a month:7.5\%,Once a week:7.5\%,Several times a week:42.5\%)} (b) - Design Proficiency: {Advanced:15\%,Beginner:5\%,Expert:7.5\%,Intermediate:57.5\%,Novice:15\%}(c) - Design Tasks performed: {Designing or creating visual content:100\%, Editing photos or videos:77.5\%, Writing reports or essays:65\%, Conducting research:62.5\%, Creating presentations or slideshows:62.5\%, Analyzing data or spreadsheets:55\%, Managing social media accounts:50\%, Coding or programming:10\%, None of the above:2.5\%} (d) - Design tools used in the past month: {Canva:90\%, PowerPoint:57.5\%, Photoshop:55\%, Google Slides:45\%, Adobe Express:32.5\%, Illustrator:20\%, Keynote:17.5\%, Prezi:15\%, Inkscape:7.5\%, Other:7.5\%, Sketch:5\%}
    
    }
    
\end{figure*}

\changed{\subsection{Study 1 User Demographics}}
\changed{Figure~\ref{fig:demographics} provides a summary of the demographic information from the survey in Section \ref{sec:study_1_screening} for the 40 users selected for the study. 70\% of the users used design tools at least once a week. Similarly  70\% of the users identified as at least Intermediate level designers with the majority (57.5\%)  identifying as intermediate level. This is typical for design tools such as Canva or Express (a slection criteria) which serve broad user populations that have design needs beyond professional designers. The most popular tools for the users were Canva (90\%), Powerpoint (57.5\%), Photoshop (55\%), Google Slides (45\%), and Adobe Express(32.5\%). All users use the tools to design or create visual content (a selection criteria), but also use these tools to edit media (77.5\%), write reports (65\%), and create presentations (62.5\%). 
}

\subsection{Survey Instructions}
\label{sec:survey_protocol}
After obtaining participation consent, survey respondents were provided the following instructions for the task:

\begin{quote}
We are investigating how different design elements affect the visual appeal and readability of text. Your participation will help us understand better design practices that ensure text is accessible. 
\begin{enumerate}
    \item You will be presented with a series of incomplete designs: a color background and one text statement (i.e., "The quick brown fox jumps over the lazy dog").
    \item For each design, you will be asked to rate the \textbf{visual appeal} and \textbf{readability} (how easy it is to read the text).
    \item When considering your responses on visual appeal, please focus on the visual appeal of the text color, background color and design elements. \textbf{The text content, font size and style will NOT CHANGE throughout the study.}
\end{enumerate}
\end{quote}

\subsection{Results \& Statistics}

%%%%%% Mean CI 
\begin{table*}[]
\centering
\caption{Mean and 95\% CI for individual Likert ratings.}
\label{table:likert_means}
\Description{Statistical data for Likert ratings across two dimensions: readability and visual appeal. The table is split into two main sections, with each section showing results for three background types (gradient, image, and texture) across seven text interventions (bckg, bckg_op, bk, out, par_op, par_out, and wh). For each intervention, the table reports the mean score, standard error (SE), and 95\% confidence interval (CI).
}
\begin{tabular}{llllllllll}
\multicolumn{5}{c}{\cellcolor[HTML]{C0C0C0}{\color[HTML]{333333} \textbf{readability}}}                                      & {\color[HTML]{FFFFFF} }                               & \multicolumn{4}{c}{\cellcolor[HTML]{C0C0C0}\textbf{visual appeal}}          \\
                                    & \textbf{intervention} & \textbf{mean} & \textbf{SE} & \multicolumn{1}{l|}{\textbf{CI}} & \multicolumn{1}{l|}{{\color[HTML]{FFFFFF} \textbf{}}} & \textbf{intervention}           & \textbf{mean} & \textbf{SE} & \textbf{CI} \\ \cline{1-5} \cline{7-10} 
                                    & bckg                  & 4.89          & 0.0532      & \multicolumn{1}{l|}{4.79, 5.00}  & \multicolumn{1}{l|}{{\color[HTML]{FFFFFF} }}          & bckg                            & 3.23          & 0.221       & 2.81, 3.66  \\
                                    & bckg\_op              & 4.82          & 0.0729      & \multicolumn{1}{l|}{4.68, 4.96}  & \multicolumn{1}{l|}{{\color[HTML]{FFFFFF} }}          & bckg\_op                        & 3.41          & 0.207       & 3.01, 3.82  \\
                                    & bk                    & 4.89          & 0.0521      & \multicolumn{1}{l|}{4.79, 5.00}  & \multicolumn{1}{l|}{{\color[HTML]{FFFFFF} }}          & bk                              & 3.80          & 0.191       & 3.42, 4.17  \\
                                    & out                   & 4.16          & 0.1621      & \multicolumn{1}{l|}{3.84, 4.47}  & \multicolumn{1}{l|}{{\color[HTML]{FFFFFF} }}          & out                             & 3.57          & 0.191       & 3.20, 3.95  \\
                                    & par\_op               & 4.17          & 0.1594      & \multicolumn{1}{l|}{3.85, 4.48}  & \multicolumn{1}{l|}{{\color[HTML]{FFFFFF} }}          & par\_op                         & 3.82          & 0.184       & 3.46, 4.18  \\
                                    & par\_out              & 4.05          & 0.1709      & \multicolumn{1}{l|}{3.71, 4.38}  & \multicolumn{1}{l|}{{\color[HTML]{FFFFFF} }}          & par\_out                        & 3.50          & 0.195       & 3.12, 3.89  \\
\multirow{-7}{*}{\textbf{gradient}} & wh                    & 3.34          & 0.2172      & \multicolumn{1}{l|}{2.91, 3.77}  & \multicolumn{1}{l|}{{\color[HTML]{FFFFFF} }}          & wh                              & 3.55          & 0.196       & 3.16, 3.93  \\ \cline{1-5} \cline{7-10} 
                                    & bckg                  & 4.81          & 0.0790      & \multicolumn{1}{l|}{4.66, 4.97}  & \multicolumn{1}{l|}{{\color[HTML]{FFFFFF} }}          & {\color[HTML]{333333} bckg}     & 2.96          & 0.228       & 2.51, 3.40  \\
                                    & bckg\_op              & 4.47          & 0.1387      & \multicolumn{1}{l|}{4.19, 4.74}  & \multicolumn{1}{l|}{}                                 & {\color[HTML]{333333} bckg\_op} & 3.61          & 0.204       & 3.20, 4.01  \\
                                    & bk                    & 2.23          & 0.2093      & \multicolumn{1}{l|}{1.82, 2.64}  & \multicolumn{1}{l|}{}                                 & {\color[HTML]{333333} bk}       & 2.35          & 0.191       & 1.98, 2.73  \\
                                    & out                   & 4.20          & 0.1677      & \multicolumn{1}{l|}{3.87, 4.52}  & \multicolumn{1}{l|}{}                                 & {\color[HTML]{333333} out}      & 3.50          & 0.206       & 3.10, 3.90  \\
                                    & par\_op               & 3.87          & 0.1994      & \multicolumn{1}{l|}{3.48, 4.26}  & \multicolumn{1}{l|}{}                                 & {\color[HTML]{333333} par\_op}  & 3.29          & 0.218       & 2.86, 3.72  \\
                                    & par\_out              & 3.75          & 0.2056      & \multicolumn{1}{l|}{3.35, 4.15}  & \multicolumn{1}{l|}{}                                 & {\color[HTML]{333333} par\_out} & 3.31          & 0.222       & 2.88, 3.75  \\
\multirow{-7}{*}{\textbf{image}}    & wh                    & 4.63          & 0.1144      & \multicolumn{1}{l|}{4.40, 4.85}  & \multicolumn{1}{l|}{}                                 & {\color[HTML]{333333} wh}       & 3.80          & 0.189       & 3.43, 4.17  \\ \cline{1-5} \cline{7-10} 
                                    & bckg                  & 4.67          & 0.1085      & \multicolumn{1}{l|}{4.46, 4.89}  & \multicolumn{1}{l|}{}                                 & {\color[HTML]{333333} bckg}     & 3.83          & 0.186       & 3.47, 4.20  \\
                                    & bckg\_op              & 4.27          & 0.1580      & \multicolumn{1}{l|}{3.96, 4.58}  & \multicolumn{1}{l|}{}                                 & {\color[HTML]{333333} bckg\_op} & 3.54          & 0.199       & 3.15, 3.93  \\
                                    & bk                    & 1.62          & 0.1646      & \multicolumn{1}{l|}{1.30, 1.94}  & \multicolumn{1}{l|}{}                                 & {\color[HTML]{333333} bk}       & 2.27          & 0.191       & 1.90, 2.65  \\
                                    & out                   & 3.58          & 0.2075      & \multicolumn{1}{l|}{3.18, 3.99}  & \multicolumn{1}{l|}{}                                 & {\color[HTML]{333333} out}      & 3.32          & 0.201       & 2.93, 3.72  \\
                                    & par\_op               & 3.69          & 0.1988      & \multicolumn{1}{l|}{3.30, 4.08}  & \multicolumn{1}{l|}{}                                 & {\color[HTML]{333333} par\_op}  & 3.42          & 0.193       & 3.04, 3.80  \\
                                    & par\_out              & 3.40          & 0.2212      & \multicolumn{1}{l|}{2.96, 3.83}  & \multicolumn{1}{l|}{}                                 & {\color[HTML]{333333} par\_out} & 2.82          & 0.209       & 2.41, 3.23  \\
\multirow{-7}{*}{\textbf{texture}}  & wh                    & 3.74          & 0.1890      & \multicolumn{1}{l|}{3.37, 4.11}  & \multicolumn{1}{l|}{}                                 & {\color[HTML]{333333} wh}       & 3.44          & 0.201       & 3.04, 3.83  \\ \cline{1-5} \cline{7-10} 
\end{tabular}
\end{table*}

%%%%%%%% Readability pairwise
\begin{table*}[]
\centering
\caption{\textbf{Readability} ratings pairwise comparison between interventions. Upper triangle shows p-values with Tukey's adjustment for multiple comparisons. Diagonals shows estimates. Lower triangle shows estimate comparisons earlier vs. later.}
\label{table:pairwise_read}
\Description{Pairwise comparison matrix of readability ratings between different text interventions (bckg, bckg_op, bk, out, par_op, par_out, wh) across gradient, image, and texture backgrounds. The matrix displays p-values with Tukey's adjustment for multiple comparisons in the upper triangle (with significant values in purple), mean ratings for each intervention on the diagonal in brackets, and estimate differences between interventions (earlier vs. later) in the lower triangle.}
% For gradient backgrounds, bckg, bckg_op, and bk performed similarly well (means ≈ 4.8) and significantly better than other interventions (p < 0.001). With image backgrounds, bckg (4.81) and wh (4.63) showed strong performance while bk performed poorly (2.23), with many significant pairwise differences. For texture backgrounds, bckg led performance (4.67) while bk performed worst (1.62), showing significant differences from other interventions (p < 0.001), with remaining interventions showing more comparable performance to each other.
\begin{tabular}{lrrrrrrr}
\rowcolor[HTML]{C0C0C0} 
\multicolumn{8}{c}{\cellcolor[HTML]{C0C0C0}\textbf{gradient}} \\
\rowcolor[HTML]{EFEFEF} 
 &
  \multicolumn{1}{l}{\cellcolor[HTML]{EFEFEF}{\color[HTML]{000000} \textbf{bckg}}} &
  \multicolumn{1}{l}{\cellcolor[HTML]{EFEFEF}{\color[HTML]{000000} \textbf{bckg\_op}}} &
  \multicolumn{1}{l}{\cellcolor[HTML]{EFEFEF}{\color[HTML]{000000} \textbf{bk}}} &
  \multicolumn{1}{l}{\cellcolor[HTML]{EFEFEF}{\color[HTML]{000000} \textbf{out}}} &
  \multicolumn{1}{l}{\cellcolor[HTML]{EFEFEF}{\color[HTML]{000000} \textbf{par\_op}}} &
  \multicolumn{1}{l}{\cellcolor[HTML]{EFEFEF}{\color[HTML]{000000} \textbf{par\_out}}} &
  \multicolumn{1}{l}{\cellcolor[HTML]{EFEFEF}{\color[HTML]{000000} \textbf{wh}}} \\
\rowcolor[HTML]{FFFFFF} 
\cellcolor[HTML]{EFEFEF}\textbf{bckg} &
  {[}4.89{]} &
  0.9790 &
  1.0000 &
  {\color[HTML]{6665CD} \textless{}.0001} &
  {\color[HTML]{6665CD} \textless{}.0001} &
  {\color[HTML]{6665CD} \textless{}.0001} &
  {\color[HTML]{6665CD} \textless{}.0001} \\
\rowcolor[HTML]{FFFFFF} 
\cellcolor[HTML]{EFEFEF}\textbf{bckg\_op} &
  0.06969 &
  {[}4.82{]} &
  0.9747 &
  {\color[HTML]{9698ED} 0.0006} &
  {\color[HTML]{9698ED} 0.0006} &
  {\color[HTML]{6665CD} \textless{}.0001} &
  {\color[HTML]{6665CD} \textless{}.0001} \\
\rowcolor[HTML]{FFFFFF} 
\cellcolor[HTML]{EFEFEF}\textbf{bk} &
  -0.00208 &
  -0.07176 &
  {[}4.89{]} &
  {\color[HTML]{6665CD} \textless{}.0001} &
  {\color[HTML]{6665CD} \textless{}.0001} &
  {\color[HTML]{6665CD} \textless{}.0001} &
  {\color[HTML]{6665CD} \textless{}.0001} \\
\rowcolor[HTML]{FFFFFF} 
\cellcolor[HTML]{EFEFEF}\textbf{out} &
  0.73475 &
  0.66506 &
  0.73682 &
  {[}4.16{]} &
  1.0000 &
  0.9976 &
  {\color[HTML]{9698ED} 0.0050} \\
\rowcolor[HTML]{FFFFFF} 
\cellcolor[HTML]{EFEFEF}\textbf{par\_op} &
  0.72452 &
  0.65483 &
  0.72659 &
  -0.01023 &
  {[}4.17{]} &
  0.9959 &
  {\color[HTML]{9698ED} 0.0037} \\
\rowcolor[HTML]{FFFFFF} 
\cellcolor[HTML]{EFEFEF}\textbf{par\_out} &
  0.84517 &
  0.77548 &
  0.84724 &
  0.11042 &
  0.12065 &
  {[}4.05{]} &
  {\color[HTML]{9698ED} 0.0312} \\
\rowcolor[HTML]{FFFFFF} 
\cellcolor[HTML]{EFEFEF}\textbf{wh} &
  1.55069 &
  1.48100 &
  1.55276 &
  0.81594 &
  0.82617 &
  0.70552 &
  {[}3.34{]} \\
 &
  \multicolumn{1}{l}{} &
  \multicolumn{1}{l}{} &
  \multicolumn{1}{l}{} &
  \multicolumn{1}{l}{} &
  \multicolumn{1}{l}{} &
  \multicolumn{1}{l}{} &
  \multicolumn{1}{l}{} \\
\rowcolor[HTML]{C0C0C0} 
\multicolumn{8}{c}{\cellcolor[HTML]{C0C0C0}\textbf{image}} \\
 &
  \multicolumn{1}{l}{\cellcolor[HTML]{EFEFEF}\textbf{bckg}} &
  \multicolumn{1}{l}{\cellcolor[HTML]{EFEFEF}\textbf{bckg\_op}} &
  \multicolumn{1}{l}{\cellcolor[HTML]{EFEFEF}\textbf{bk}} &
  \multicolumn{1}{l}{\cellcolor[HTML]{EFEFEF}\textbf{out}} &
  \multicolumn{1}{l}{\cellcolor[HTML]{EFEFEF}\textbf{par\_op}} &
  \multicolumn{1}{l}{\cellcolor[HTML]{EFEFEF}\textbf{par\_out}} &
  \multicolumn{1}{l}{\cellcolor[HTML]{EFEFEF}\textbf{wh}} \\
\rowcolor[HTML]{FFFFFF} 
\cellcolor[HTML]{EFEFEF}\textbf{bckg} &
  {[}4.81{]} &
  0.1939 &
  {\color[HTML]{6665CD} \textless{}.0001} &
  {\color[HTML]{9698ED} 0.0046} &
  {\color[HTML]{6665CD} \textless{}.0001} &
  {\color[HTML]{6665CD} \textless{}.0001} &
  0.7614 \\
\rowcolor[HTML]{FFFFFF} 
\cellcolor[HTML]{EFEFEF}\textbf{bckg\_op} &
  0.344 &
  {[}4.47{]} &
  {\color[HTML]{6665CD} \textless{}.0001} &
  0.7660 &
  0.0624 &
  {\color[HTML]{9698ED} 0.0135} &
  0.9411 \\
\rowcolor[HTML]{FFFFFF} 
\cellcolor[HTML]{EFEFEF}\textbf{bk} &
  2.585 &
  2.241 &
  {[}2.23{]} &
  {\color[HTML]{6665CD} \textless{}.0001} &
  {\color[HTML]{6665CD} \textless{}.0001} &
  {\color[HTML]{6665CD} \textless{}.0001} &
  {\color[HTML]{6665CD} \textless{}.0001} \\
\rowcolor[HTML]{FFFFFF} 
\cellcolor[HTML]{EFEFEF}\textbf{out} &
  0.615 &
  0.271 &
  -1.970 &
  {[}4.20{]} &
  0.7526 &
  0.4243 &
  0.1695 \\
\rowcolor[HTML]{FFFFFF} 
\cellcolor[HTML]{EFEFEF}\textbf{par\_op} &
  0.943 &
  0.599 &
  -1.642 &
  0.328 &
  {[}3.87{]} &
  0.9991 &
  {\color[HTML]{9698ED} 0.0030} \\
\rowcolor[HTML]{FFFFFF} 
\cellcolor[HTML]{EFEFEF}\textbf{par\_out} &
  1.060 &
  0.716 &
  -1.525 &
  0.445 &
  0.117 &
  {[}3.75{]} &
  {\color[HTML]{9698ED} 0.0004} \\
\rowcolor[HTML]{FFFFFF} 
\cellcolor[HTML]{EFEFEF}\textbf{wh} &
  0.182 &
  -0.162 &
  -2.403 &
  -0.433 &
  -0.761 &
  -0.878 &
  {[}4.63{]} \\
 &
  \multicolumn{1}{l}{} &
  \multicolumn{1}{l}{} &
  \multicolumn{1}{l}{} &
  \multicolumn{1}{l}{} &
  \multicolumn{1}{l}{} &
  \multicolumn{1}{l}{} &
  \multicolumn{1}{l}{} \\
\rowcolor[HTML]{C0C0C0} 
\multicolumn{8}{c}{\cellcolor[HTML]{C0C0C0}\textbf{texture}} \\
 &
  \multicolumn{1}{l}{\cellcolor[HTML]{EFEFEF}\textbf{bckg}} &
  \multicolumn{1}{l}{\cellcolor[HTML]{EFEFEF}\textbf{bckg\_op}} &
  \multicolumn{1}{l}{\cellcolor[HTML]{EFEFEF}\textbf{bk}} &
  \multicolumn{1}{l}{\cellcolor[HTML]{EFEFEF}\textbf{out}} &
  \multicolumn{1}{l}{\cellcolor[HTML]{EFEFEF}\textbf{par\_op}} &
  \multicolumn{1}{l}{\cellcolor[HTML]{EFEFEF}\textbf{par\_out}} &
  \multicolumn{1}{l}{\cellcolor[HTML]{EFEFEF}\textbf{wh}} \\
\cellcolor[HTML]{EFEFEF}\textbf{bckg} &
  \multicolumn{1}{l}{{[}4.67{]}} &
  \multicolumn{1}{l}{0.1776} &
  \multicolumn{1}{l}{{\color[HTML]{6665CD} \textless{}.0001}} &
  \multicolumn{1}{l}{\cellcolor[HTML]{FFFFFF}{\color[HTML]{6665CD} \textless{}.0001}} &
  \multicolumn{1}{l}{\cellcolor[HTML]{FFFFFF}{\color[HTML]{6665CD} \textless{}.0001}} &
  \multicolumn{1}{l}{\cellcolor[HTML]{FFFFFF}{\color[HTML]{6665CD} \textless{}.0001}} &
  \multicolumn{1}{l}{\cellcolor[HTML]{FFFFFF}{\color[HTML]{6665CD} \textless{}.0001}} \\
\cellcolor[HTML]{EFEFEF}\textbf{bckg\_op} &
  \multicolumn{1}{l}{0.407} &
  \multicolumn{1}{l}{{[}4.27{]}} &
  \multicolumn{1}{l}{{\color[HTML]{6665CD} \textless{}.0001}} &
  \multicolumn{1}{l}{\cellcolor[HTML]{FFFFFF}{\color[HTML]{9698ED} 0.0291}} &
  \multicolumn{1}{l}{\cellcolor[HTML]{FFFFFF}0.0919} &
  \multicolumn{1}{l}{\cellcolor[HTML]{FFFFFF}{\color[HTML]{9698ED} 0.0027}} &
  \multicolumn{1}{l}{\cellcolor[HTML]{FFFFFF}0.1239} \\
\cellcolor[HTML]{EFEFEF}\textbf{bk} &
  \multicolumn{1}{l}{3.053} &
  \multicolumn{1}{l}{2.647} &
  \multicolumn{1}{l}{{[}1.62{]}} &
  \multicolumn{1}{l}{\cellcolor[HTML]{FFFFFF}{\color[HTML]{6665CD} \textless{}.0001}} &
  \multicolumn{1}{l}{\cellcolor[HTML]{FFFFFF}{\color[HTML]{6665CD} \textless{}.0001}} &
  \multicolumn{1}{l}{\cellcolor[HTML]{FFFFFF}{\color[HTML]{6665CD} \textless{}.0001}} &
  \multicolumn{1}{l}{\cellcolor[HTML]{FFFFFF}{\color[HTML]{6665CD} \textless{}.0001}} \\
\cellcolor[HTML]{EFEFEF}\textbf{out} &
  \multicolumn{1}{l}{1.091} &
  \multicolumn{1}{l}{0.684} &
  \multicolumn{1}{l}{-1.963} &
  \multicolumn{1}{l}{\cellcolor[HTML]{FFFFFF}{[}3.58{]}} &
  \multicolumn{1}{l}{\cellcolor[HTML]{FFFFFF}0.9994} &
  \multicolumn{1}{l}{\cellcolor[HTML]{FFFFFF}0.9890} &
  \multicolumn{1}{l}{\cellcolor[HTML]{FFFFFF}0.9942} \\
\cellcolor[HTML]{EFEFEF}\textbf{par\_op} &
  \multicolumn{1}{l}{0.984} &
  \multicolumn{1}{l}{0.577} &
  \multicolumn{1}{l}{-2.069} &
  \multicolumn{1}{l}{\cellcolor[HTML]{FFFFFF}-0.106} &
  \multicolumn{1}{l}{\cellcolor[HTML]{FFFFFF}{[}3.69{]}} &
  \multicolumn{1}{l}{\cellcolor[HTML]{FFFFFF}0.8927} &
  \multicolumn{1}{l}{\cellcolor[HTML]{FFFFFF}1.0000} \\
\cellcolor[HTML]{EFEFEF}\textbf{par\_out} &
  \multicolumn{1}{l}{1.279} &
  \multicolumn{1}{l}{0.872} &
  \multicolumn{1}{l}{-1.775} &
  \multicolumn{1}{l}{\cellcolor[HTML]{FFFFFF}0.188} &
  \multicolumn{1}{l}{\cellcolor[HTML]{FFFFFF}0.295} &
  \multicolumn{1}{l}{\cellcolor[HTML]{FFFFFF}{[}3.40{]}} &
  \multicolumn{1}{l}{\cellcolor[HTML]{FFFFFF}0.7826} \\
\cellcolor[HTML]{EFEFEF}\textbf{wh} &
  \multicolumn{1}{l}{0.937} &
  \multicolumn{1}{l}{0.530} &
  \multicolumn{1}{l}{-2.116} &
  \multicolumn{1}{l}{-0.153} &
  \multicolumn{1}{l}{-0.047} &
  \multicolumn{1}{l}{-0.342} &
  \multicolumn{1}{l}{{[}3.74{]}}
\end{tabular}
\end{table*}

%%%%%%%% Aes pairwise

\begin{table*}[]
\centering
\caption{\textbf{Visual appeal} ratings pairwise comparison between interventions. Upper triangle shows p-values with Tukey's adjustment for multiple comparisons. Diagonals shows estimates. Lower triangle shows estimate comparisons earlier vs. later.}
\label{table:pairwise_aes}
\Description{Pairwise comparison matrix examining visual appeal ratings between different text interventions (bckg, bckg_op, bk, out, par_op, par_out, wh) across gradient, image, and texture backgrounds. The matrix shows p-values with Tukey's adjustment in the upper triangle (significant values in purple), mean ratings on the diagonal in brackets, and estimate differences between interventions in the lower triangle.}
\begin{tabular}{lrrrrrrr}
\rowcolor[HTML]{C0C0C0} 
\multicolumn{8}{c}{\cellcolor[HTML]{C0C0C0}\textbf{gradient}} \\
\rowcolor[HTML]{EFEFEF} 
 &
  \multicolumn{1}{l}{\cellcolor[HTML]{EFEFEF}{\color[HTML]{000000} \textbf{bckg}}} &
  \multicolumn{1}{l}{\cellcolor[HTML]{EFEFEF}{\color[HTML]{000000} \textbf{bckg\_op}}} &
  \multicolumn{1}{l}{\cellcolor[HTML]{EFEFEF}{\color[HTML]{000000} \textbf{bk}}} &
  \multicolumn{1}{l}{\cellcolor[HTML]{EFEFEF}{\color[HTML]{000000} \textbf{out}}} &
  \multicolumn{1}{l}{\cellcolor[HTML]{EFEFEF}{\color[HTML]{000000} \textbf{par\_op}}} &
  \multicolumn{1}{l}{\cellcolor[HTML]{EFEFEF}{\color[HTML]{000000} \textbf{par\_out}}} &
  \multicolumn{1}{l}{\cellcolor[HTML]{EFEFEF}{\color[HTML]{000000} \textbf{wh}}} \\
\rowcolor[HTML]{FFFFFF} 
\cellcolor[HTML]{EFEFEF}\textbf{bckg} &
  {\color[HTML]{333333} {[}3.23{]}} &
  {\color[HTML]{333333} 0.9942} &
  {\color[HTML]{333333} 0.3045} &
  {\color[HTML]{333333} 0.8347} &
  {\color[HTML]{333333} 0.2393} &
  {\color[HTML]{333333} 0.9410} &
  {\color[HTML]{333333} 0.8862} \\
\rowcolor[HTML]{FFFFFF} 
\cellcolor[HTML]{EFEFEF}\textbf{bckg\_op} &
  {\color[HTML]{333333} -0.1789} &
  {\color[HTML]{333333} {[}3.41{]}} &
  {\color[HTML]{333333} 0.7100} &
  {\color[HTML]{333333} 0.9943} &
  {\color[HTML]{333333} 0.6320} &
  {\color[HTML]{333333} 0.9998} &
  {\color[HTML]{333333} 0.9979} \\
\rowcolor[HTML]{FFFFFF} 
\cellcolor[HTML]{EFEFEF}\textbf{bk} &
  {\color[HTML]{333333} -0.5653} &
  {\color[HTML]{333333} -0.3864} &
  {\color[HTML]{333333} {[}3.80{]}} &
  {\color[HTML]{333333} 0.9664} &
  {\color[HTML]{333333} 1.0000} &
  {\color[HTML]{333333} 0.8859} &
  {\color[HTML]{333333} 0.9472} \\
\rowcolor[HTML]{FFFFFF} 
\cellcolor[HTML]{EFEFEF}\textbf{out} &
  {\color[HTML]{333333} -0.3432} &
  {\color[HTML]{333333} -0.1643} &
  {\color[HTML]{333333} 0.2221} &
  {\color[HTML]{333333} {[}3.57{]}} &
  {\color[HTML]{333333} 0.9414} &
  {\color[HTML]{333333} 0.9999} &
  {\color[HTML]{333333} 1.0000} \\
\rowcolor[HTML]{FFFFFF} 
\cellcolor[HTML]{EFEFEF}\textbf{par\_op} &
  {\color[HTML]{333333} -0.5868} &
  {\color[HTML]{333333} -0.4079} &
  {\color[HTML]{333333} -0.0215} &
  {\color[HTML]{333333} -0.2436} &
  {\color[HTML]{333333} {[}3.82{]}} &
  {\color[HTML]{333333} 0.8324} &
  {\color[HTML]{333333} 0.9146} \\
\rowcolor[HTML]{FFFFFF} 
\cellcolor[HTML]{EFEFEF}\textbf{par\_out} &
  {\color[HTML]{333333} -0.2731} &
  {\color[HTML]{333333} -0.0942} &
  {\color[HTML]{333333} 0.2922} &
  {\color[HTML]{333333} 0.0701} &
  {\color[HTML]{333333} 0.3137} &
  {\color[HTML]{333333} {[}3.50{]}} &
  {\color[HTML]{333333} 1.0000} \\
\rowcolor[HTML]{FFFFFF} 
\cellcolor[HTML]{EFEFEF}\textbf{wh} &
  {\color[HTML]{333333} -0.3176} &
  {\color[HTML]{333333} -0.1388} &
  {\color[HTML]{333333} 0.2477} &
  {\color[HTML]{333333} 0.0256} &
  {\color[HTML]{333333} 0.2692} &
  {\color[HTML]{333333} -0.0445} &
  {\color[HTML]{333333} {[}3.55{]}} \\
 &
  \multicolumn{1}{l}{} &
  \multicolumn{1}{l}{} &
  \multicolumn{1}{l}{} &
  \multicolumn{1}{l}{} &
  \multicolumn{1}{l}{} &
  \multicolumn{1}{l}{} &
  \multicolumn{1}{l}{} \\
\rowcolor[HTML]{C0C0C0} 
\multicolumn{8}{c}{\cellcolor[HTML]{C0C0C0}\textbf{image}} \\
 &
  \multicolumn{1}{l}{\cellcolor[HTML]{EFEFEF}\textbf{bckg}} &
  \multicolumn{1}{l}{\cellcolor[HTML]{EFEFEF}\textbf{bckg\_op}} &
  \multicolumn{1}{l}{\cellcolor[HTML]{EFEFEF}\textbf{bk}} &
  \multicolumn{1}{l}{\cellcolor[HTML]{EFEFEF}\textbf{out}} &
  \multicolumn{1}{l}{\cellcolor[HTML]{EFEFEF}\textbf{par\_op}} &
  \multicolumn{1}{l}{\cellcolor[HTML]{EFEFEF}\textbf{par\_out}} &
  \multicolumn{1}{l}{\cellcolor[HTML]{EFEFEF}\textbf{wh}} \\
\rowcolor[HTML]{FFFFFF} 
\cellcolor[HTML]{EFEFEF}\textbf{bckg} &
  {\color[HTML]{333333} {[}2.96{]}} &
  {\color[HTML]{333333} 0.2091} &
  {\color[HTML]{333333} 0.2435} &
  {\color[HTML]{333333} 0.4216} &
  {\color[HTML]{333333} 0.9008} &
  {\color[HTML]{333333} 0.8717} &
  {\color[HTML]{9698ED} 0.0242} \\
\rowcolor[HTML]{FFFFFF} 
\cellcolor[HTML]{EFEFEF}\textbf{bckg\_op} &
  {\color[HTML]{333333} -0.647} &
  {\color[HTML]{333333} {[}3.61{]}} &
  {\color[HTML]{6665CD} \textless{}.0001} &
  {\color[HTML]{333333} 0.9996} &
  {\color[HTML]{333333} 0.9004} &
  {\color[HTML]{333333} 0.9335} &
  {\color[HTML]{333333} 0.9870} \\
\rowcolor[HTML]{FFFFFF} 
\cellcolor[HTML]{EFEFEF}\textbf{bk} &
  {\color[HTML]{333333} 0.605} &
  {\color[HTML]{333333} 1.252} &
  {\color[HTML]{333333} {[}2.35{]}} &
  {\color[HTML]{6665CD} \textless{}.0001} &
  {\color[HTML]{9698ED} 0.0046} &
  {\color[HTML]{9698ED} 0.0038} &
  {\color[HTML]{6665CD} \textless{}.0001} \\
\rowcolor[HTML]{FFFFFF} 
\cellcolor[HTML]{EFEFEF}\textbf{out} &
  {\color[HTML]{333333} -0.542} &
  {\color[HTML]{333333} 0.104} &
  {\color[HTML]{333333} -1.148} &
  {\color[HTML]{333333} {[}3.50{]}} &
  {\color[HTML]{333333} 0.9860} &
  {\color[HTML]{333333} 0.9931} &
  {\color[HTML]{333333} 0.8938} \\
\rowcolor[HTML]{FFFFFF} 
\cellcolor[HTML]{EFEFEF}\textbf{par\_op} &
  {\color[HTML]{333333} -0.333} &
  {\color[HTML]{333333} 0.314} &
  {\color[HTML]{333333} -0.938} &
  {\color[HTML]{333333} 0.210} &
  {\color[HTML]{333333} {[}3.29{]}} &
  {\color[HTML]{333333} 1.0000} &
  {\color[HTML]{333333} 0.4277} \\
\rowcolor[HTML]{FFFFFF} 
\cellcolor[HTML]{EFEFEF}\textbf{par\_out} &
  {\color[HTML]{333333} -0.357} &
  {\color[HTML]{333333} 0.290} &
  {\color[HTML]{333333} -0.962} &
  {\color[HTML]{333333} 0.186} &
  {\color[HTML]{333333} -0.024} &
  {\color[HTML]{333333} {[}3.31{]}} &
  {\color[HTML]{333333} 0.5060} \\
\rowcolor[HTML]{FFFFFF} 
\cellcolor[HTML]{EFEFEF}\textbf{wh} &
  {\color[HTML]{333333} -0.838} &
  {\color[HTML]{333333} -0.191} &
  {\color[HTML]{333333} -1.443} &
  {\color[HTML]{333333} -0.296} &
  {\color[HTML]{333333} -0.505} &
  {\color[HTML]{333333} -0.481} &
  {\color[HTML]{333333} {[}3.80{]}} \\
 &
  \multicolumn{1}{l}{} &
  \multicolumn{1}{l}{} &
  \multicolumn{1}{l}{} &
  \multicolumn{1}{l}{} &
  \multicolumn{1}{l}{} &
  \multicolumn{1}{l}{} &
  \multicolumn{1}{l}{} \\
\rowcolor[HTML]{C0C0C0} 
\multicolumn{8}{c}{\cellcolor[HTML]{C0C0C0}\textbf{texture}} \\
 &
  \multicolumn{1}{l}{\cellcolor[HTML]{EFEFEF}\textbf{bckg}} &
  \multicolumn{1}{l}{\cellcolor[HTML]{EFEFEF}\textbf{bckg\_op}} &
  \multicolumn{1}{l}{\cellcolor[HTML]{EFEFEF}\textbf{bk}} &
  \multicolumn{1}{l}{\cellcolor[HTML]{EFEFEF}\textbf{out}} &
  \multicolumn{1}{l}{\cellcolor[HTML]{EFEFEF}\textbf{par\_op}} &
  \multicolumn{1}{l}{\cellcolor[HTML]{EFEFEF}\textbf{par\_out}} &
  \multicolumn{1}{l}{\cellcolor[HTML]{EFEFEF}\textbf{wh}} \\
\rowcolor[HTML]{FFFFFF} 
\cellcolor[HTML]{EFEFEF}\textbf{bckg} &
  {\color[HTML]{333333} {[}3.83{]}} &
  {\color[HTML]{333333} 0.8845} &
  {\color[HTML]{6665CD} \textless{}.0001} &
  {\color[HTML]{333333} 0.3357} &
  {\color[HTML]{333333} 0.5718} &
  \cellcolor[HTML]{FFFFFF}{\color[HTML]{9698ED} 0.0008} &
  {\color[HTML]{333333} 0.6468} \\
\rowcolor[HTML]{FFFFFF} 
\cellcolor[HTML]{EFEFEF}\textbf{bckg\_op} &
  {\color[HTML]{333333} 0.2936} &
  {\color[HTML]{333333} {[}3.54{]}} &
  {\color[HTML]{6665CD} \textless{}.0001} &
  {\color[HTML]{333333} 0.9756} &
  {\color[HTML]{333333} 0.9989} &
  {\color[HTML]{333333} 0.0688} &
  {\color[HTML]{333333} 0.9996} \\
\rowcolor[HTML]{FFFFFF} 
\cellcolor[HTML]{EFEFEF}\textbf{bk} &
  {\color[HTML]{333333} 1.5605} &
  {\color[HTML]{333333} 1.2669} &
  {\color[HTML]{333333} {[}2.27{]}} &
  {\color[HTML]{9698ED} 0.0003} &
  {\color[HTML]{6665CD} \textless{}.0001} &
  {\color[HTML]{333333} 0.2866} &
  {\color[HTML]{6665CD} \textless{}.0001} \\
\rowcolor[HTML]{FFFFFF} 
\cellcolor[HTML]{EFEFEF}\textbf{out} &
  {\color[HTML]{333333} 0.5110} &
  {\color[HTML]{333333} 0.2174} &
  {\color[HTML]{333333} -1.0495} &
  {\color[HTML]{333333} {[}3.32{]}} &
  {\color[HTML]{333333} 0.9997} &
  {\color[HTML]{333333} 0.4322} &
  {\color[HTML]{333333} 0.9993} \\
\rowcolor[HTML]{FFFFFF} 
\cellcolor[HTML]{EFEFEF}\textbf{par\_op} &
  {\color[HTML]{333333} 0.4128} &
  {\color[HTML]{333333} 0.1192} &
  {\color[HTML]{333333} -1.1477} &
  {\color[HTML]{333333} -0.0982} &
  {\color[HTML]{333333} {[}3.42{]}} &
  {\color[HTML]{333333} 0.1915} &
  {\color[HTML]{333333} 1.0000} \\
\rowcolor[HTML]{FFFFFF} 
\cellcolor[HTML]{EFEFEF}\textbf{par\_out} &
  {\color[HTML]{333333} 1.0102} &
  {\color[HTML]{333333} 0.7166} &
  {\color[HTML]{333333} -0.5503} &
  {\color[HTML]{333333} 0.4992} &
  {\color[HTML]{333333} 0.5974} &
  {\color[HTML]{333333} {[}2.82{]}} &
  {\color[HTML]{333333} 0.1892} \\
\rowcolor[HTML]{FFFFFF} 
\cellcolor[HTML]{EFEFEF}\textbf{wh} &
  {\color[HTML]{333333} 0.3973} &
  {\color[HTML]{333333} 0.1037} &
  {\color[HTML]{333333} -1.1632} &
  {\color[HTML]{333333} -0.1137} &
  {\color[HTML]{333333} -0.0155} &
  {\color[HTML]{333333} -0.6129} &
  {\color[HTML]{333333} {[}3.44{]}}
\end{tabular}
\end{table*}

Table~\ref{table:likert_means} shows the readability and visual appeal Likert rating means and 95\% CI for each intervention. Table~\ref{table:pairwise_read} shows the readability pairwise comparisons p-values and Table~\ref{table:pairwise_aes} shows visual appeal pairwise comparisons.

\section{Study 2: Workflow Experience Evaluation}

\subsection{Templates}
\begin{figure*}
    \centering
    \includegraphics[width=0.85\textwidth, 
        alt={Two sets of background examples used in the design evaluation study. The first set, labeled 'texture', includes a geometric abstract pattern with blue, yellow, and red shapes, and a retro-style image with pink donuts on a red and white sunburst background. The second set, labeled 'image', shows two photographs: one featuring models in clothing against a dark setting, and another showing a snowy mountain landscape with a macro view of cheese blocks. These backgrounds were used for participants to complete their design tasks during the experience evaluation.}]{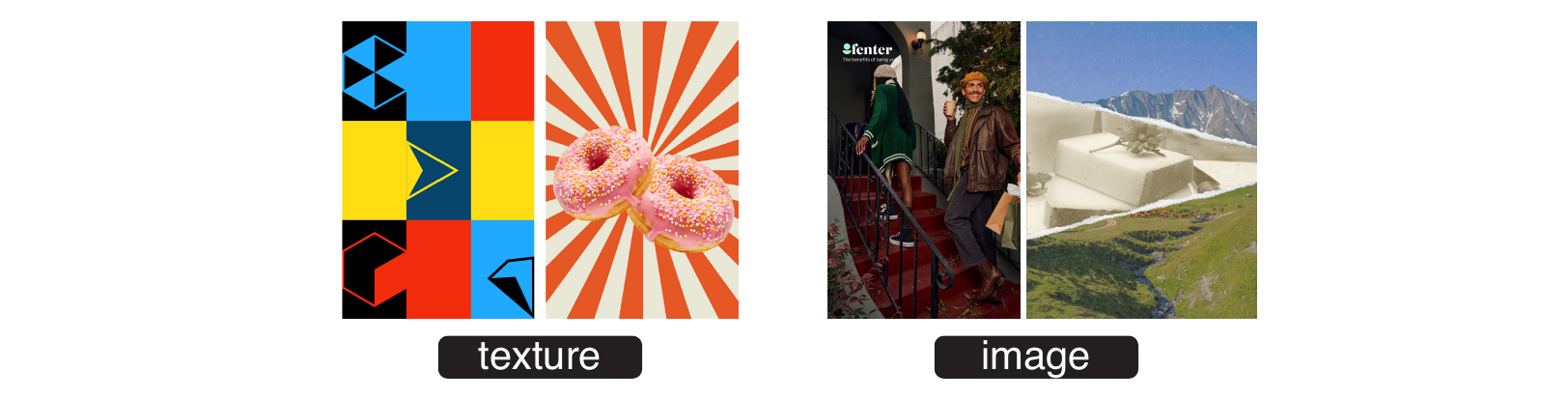}
    \caption{In the experience evaluation, participants completed two design tasks using either a texture or image background.}
    \label{fig:experience_templates}
    \Description{Two sets of background examples used in the design evaluation study. The first set, labeled 'texture', includes a geometric abstract pattern with blue, yellow, and red shapes, and a retro-style image with pink donuts on a red and white sunburst background. The second set, labeled 'image', shows two photographs: one featuring models in clothing against a dark setting, and another showing a snowy mountain landscape with a macro view of cheese blocks. These backgrounds were used for participants to complete their design tasks during the experience evaluation.}
\end{figure*}

Participants completed two design tasks during the study. They were provided with text and a design background (either an image or texture) and were instructed to create a draft of their design. The backgrounds used are shown in Figure~\ref{fig:experience_templates}. 

\begin{figure*}
    \centering
    \includegraphics[width=\textwidth, 
        alt={Adobe Express interface with an integrated color contrast checker panel used as the baseline tool in the study. The main interface displays a promotional image for cheese making with mountains in the background. To the right, a 'Contrast Checker' panel is shown with three key features annotated with arrows: 1) User-selectable foreground/background color fields with hex codes (#00087a and #c7e3ea shown), 2) A contrast ratio display showing '11.86:1', and 3) WCAG compliance indicators showing 'Pass' status for Regular Text, Large Text, and Graphical Objects under both AA and AAA standards.}]{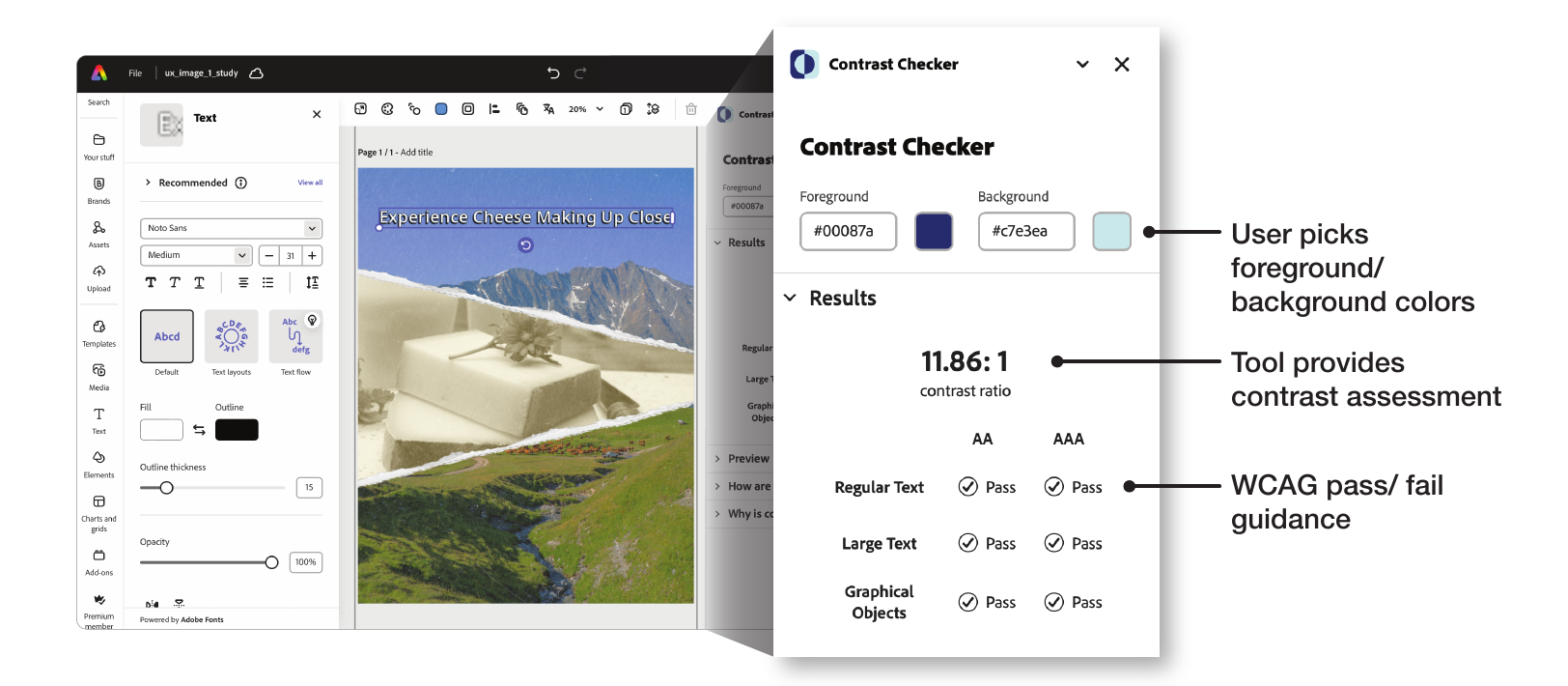}
    \caption{The Adobe Express with a color contrast checker panel was used as a baseline during the experience evaluation (Section~\ref{sec:study_2_experience}).}
    \label{fig:baseline}
    \Description{Adobe Express interface with an integrated color contrast checker panel used as the baseline tool in the study. The main interface displays a promotional image for cheese making with mountains in the background. To the right, a 'Contrast Checker' panel is shown with three key features annotated with arrows: 1) User-selectable foreground/background color fields with hex codes (#00087a and #c7e3ea shown), 2) A contrast ratio display showing '11.86:1', and 3) WCAG compliance indicators showing 'Pass' status for Regular Text, Large Text, and Graphical Objects under both AA and AAA standards.}
\end{figure*}

\subsection{Baseline Tool}
As a baseline comparison, participants used Adobe Express and a typical color contrast checker add-on which required participants to probe for specific foreground/background colors (Figure~\ref{fig:baseline}). The tool provided users with a color contrast value and a pass/fail assessment based on WCAG guidelines.

\subsection{Semi-Structured Interview Guide}
\label{sec:interview_guide}

We used a think aloud protocol while users completed a design task using \system{} and the baseline tool. In addition, we conducted a semi-structured interview after using each tool and at the end of the entire session to collect qualitative feedback and user impressions on both tools. The subsections below describe the questions which guided the study moderator's conversation.

\subsubsection{Current Workflows \& Challenges \textit{(beginning of session)}}
\begin{enumerate}
    \item Can you describe your typical design process? Walk me through a past project.
    \item How do you usually approach accessibility in your designs? 
    \item What challenges do you encounter when trying to make your designs accessible? Show me an example.
\end{enumerate}

\subsubsection{Tool Feedback \textit{(after each tool)}}
\begin{enumerate}
    \item What are your overall impressions of the tool you just tried?
    \item Were there any features that you found particularly useful or confusing/unnecessary?
    \item Are you satisfied with your final design?
    \item Are you confident your design meets color contrast accessibility standards?
\end{enumerate}

\subsubsection{Workflow Integration \textit{(end of session)}}
\begin{enumerate}
    \item How do the tools you used today compare to your current workflow managing color accessibility?
    \item How do you envision incorporating this tool into your existing workflow, if at all?
    \item What changes or improvements would make this tool more useful in your work?
    \item Are there any additional features you think would enhance the tool's effectiveness?
    \item After today’s session, has your perspective on designing for color accessibility changed? If so, how?
\end{enumerate}
%TC:endignore

\end{document}